\documentclass[12pt]{article}
\usepackage{graphicx}
\usepackage{epsfig}
\usepackage{amsmath}
\begin{document}

\def\a{\alpha}
\def\b{\beta}
\def\c{\varepsilon}
\def\d{\delta}
\def\e{\epsilon}
\def\f{\phi}
\def\g{\gamma}
\def\h{\theta}
\def\k{\kappa}
\def\l{\lambda}
\def\m{\mu}
\def\n{\nu}
\def\p{\psi}
\def\q{\partial}
\def\r{\rho}
\def\s{\sigma}
\def\t{\tau}
\def\u{\upsilon}
\def\v{\varphi}
\def\w{\omega}
\def\x{\xi}
\def\y{\eta}
\def\z{\zeta}
\def\D{\Delta}
\def\G{\Gamma}
\def\H{\Theta}
\def\L{\Lambda}
\def\F{\Phi}
\def\P{\Psi}
\def\S{\Sigma}

\def\o{\over}
\def\beq{\begin{eqnarray}}
\def\eeq{\end{eqnarray}}
\newcommand{\lsim}{\raisebox{0.6mm}{$\, <$} \hspace{-3.0mm}\raisebox{-1.5mm}{\em $\sim \,$}}
\newcommand{\gsim}{\raisebox{0.6mm}{$\, >$} \hspace{-3.0mm}\raisebox{-1.5mm}{\em $\sim \,$}}

\renewcommand{\floatpagefraction}{0.7}
\newcommand{\vev}[1]{ \left\langle {#1} \right\rangle }
\newcommand{\bra}[1]{ \langle {#1} | }
\newcommand{\ket}[1]{ | {#1} \rangle }
\newcommand{\EV}{ {\rm eV} }
\newcommand{\keV}{ {\rm keV} }
\newcommand{\MeV}{ {\rm MeV} }
\newcommand{\GeV}{ {\rm GeV} }
\newcommand{\TeV}{ {\rm TeV} }
\def\diag{\mathop{\rm diag}\nolimits}
\def\Spin{\mathop{\rm Spin}}
\def\SO{\mathop{\rm SO}}
\def\O{\mathop{\rm O}}
\def\SU{\mathop{\rm SU}}
\def\U{\mathop{\rm U}}
\def\Sp{\mathop{\rm Sp}}
\def\SL{\mathop{\rm SL}}
\def\tr{\mathop{\rm tr}}

\def\Z{\mathcal{Z}}
\def\W{\Omega}

\baselineskip 0.7cm

\begin{titlepage}

\begin{flushright}
CALT-68-2780\\
IPMU10-0045\\
UT-10-03
\end{flushright}

\vskip 1.35cm
\begin{center}
{\large \bf

Aspects of Non-minimal Gauge Mediation 

}
\vskip 1.2cm
Satoshi Shirai$^{1,2}$, Masahito Yamazaki$^{1,2,3}$ and Kazuya Yonekura$^{1,2}$
\vskip 0.4cm

{\it
$^1$Department of Physics, University of Tokyo, 
Tokyo 113-0033,
Japan\\

$^2$IPMU,
University of Tokyo, 
Chiba 277-8586, 
Japan\\

$^3$California Institute of Technology, 
CA 91125, USA
}

\vskip 1.5cm

\abstract{
A large class of non-minimal gauge mediation models, such as (semi-)direct gauge mediation, predict a hierarchy between the masses of the supersymmetric standard model gauginos and those of scalar particles. We perform a comprehensive study of these non-minimal gauge mediation models, including mass calculations in semi-direct gauge mediation, to illustrate these features, and discuss the phenomenology of the models. 
We point out that the cosmological gravitino problem places stringent constraints on mass splittings, when 
the Bino is the NLSP. However, the GUT relation of the gaugino masses is broken unlike the case of minimal gauge mediation,
and an NLSP other than the Bino (especially the gluino NLSP) becomes possible, relaxing the cosmological constraints.
We also discuss the collider signals of the models.
}

\end{center}
\end{titlepage}

\setcounter{page}{2}

\tableofcontents

\section{Introduction}

Gauge-mediated supersymmetry breaking (GMSB) \cite{Dine:1981za,mGMSB,Giudice:1998bp} is an attractive scenario for communicating supersymmetry (SUSY) breaking to the supersymmetric Standard Model (SSM). It elegantly solves the problems of Flavor-Changing Neutral Current (FCNC), which is rather problematic in conventional gravity mediation scenario. Moreover, it does not rely on physics at the ultraviolet scale (Planck scale), and can be discussed in the well-established framework of quantum field theory.

In this paper, we discuss non-minimal models of GMSB.
The minimal gauge mediation (mGMSB) \cite{mGMSB} is probably the most famous model of GMSB and still continues to be a viable possibility. However, there are also other non-minimal models of GMSB, including direct mediation and more recently semi-direct gauge mediation  \cite{Izawa:1997hu,Ibe:2007wp,Seiberg:2008qj}. As we will discuss in the next section, a large class of these non-minimal gauge mediation models share the common feature that there is a large hierarchy between the masses of SSM gauginos and those of scalar particles. 
We call such models of GMSB the ``split gauge mediation''~\footnote{The splitting discussed in this paper is of order $\lsim {\cal O}(10^3)$, as we will see.} (split GMSB).

One problem of the split GMSB is that the existence of a hierarchy is against the philosophy of naturalness, which is one of the conventional motivations for supersymmetry. However, we should be careful when referring to naturalness --- the hierarchy here is much smaller than the large hierarchy between electroweak (EW) scale and the Planck scale, and Nature may still allow such ``little hierarchy'' \footnote{
Note that even if we choose mGMSB, we still have another ``little hierarchy problem'' arising from the LEP-II mass bound for a Standard Model Higgs,
since $m_{h^0} > 114~\GeV$ implies $m_{\tilde q} \gsim 1~\TeV$.
}. There are also many theoretically motivated models of supersymmetry with split mass spectrum, such as anomaly mediation (AMSB)~\footnote{
AMSB referred to in this paper is the one in which the gravity mediation effects to the sfermions are not sequestered
and the sfermion masses are of the order of the gravitino mass.
}~\cite{AMSB}, focus point \cite{FocusPoint} and split SUSY \cite{splitSUSY} \footnote{
Readers should keep in mind, however, that there are important differences between split GMSB discussed in this paper and split SUSY. For example, we typically consider ${\cal O}(10^1-10^2)$ splittings, whereas in split SUSY the splittings are much larger. This means ${\cal O}(1)$ mixing among different flavors are often excluded by FCNC constraints in split GMSB. Moreover, the value of the $\mu$ is much larger than electroweak scale in split GMSB, if there is no fine-tuning.}. Moreover, in some extreme scenarios such as split SUSY, SUSY can be motivated only by Lightest Supersymmetric Particle (LSP) dark matter and gauge coupling unification, without any reference to the gauge hierarchy problem. There are also indications that supersymmetry is essential for the construction of consistent theories of quantum gravity, such as string theory. Even if we assume naturalness it is often unclear to what extent the splitting is allowed (e.g. whether or not we allow ${\cal O}(10)$ splittings), and different people will have different attitudes toward this problem.

In this paper, we therefore rely exclusively on phenomenological constraints on the mass splittings, instead of invoking naturalness. We assume throughout this paper that gravitino is the LSP (except for the short discussion of heavy gravitino scenario in the final section). The phenomenology depends drastically on the choice of Next-to-Lightest Supersymmetric Particle (NLSP). We will see that when the Bino is the NLSP, the split GMSB models are heavily constrained by cosmological gravitino problems, leaving the Wino and gluino as viable candidates of the NLSP. We also show that the GUT relation of the SSM gaugino masses,
$M_{\tilde B}:M_{\tilde W}:M_{\tilde g}\simeq \a_1:\a_2:\a_3$, is broken in the split GMSB, and in particular the gluino tends to be light, often becoming the NLSP. This leads to a characteristic mass spectrum, which is different from that of typical GMSB models.
We also discuss the collider signals of the models, taking the cosmological constraints into account.

This paper is organized as follows. In section \ref{sec:2} and section \ref{sec:3}, we give detailed discussion of mass splittings, based both on general considerations (section \ref{sec:2}) and specific examples (section \ref{sec:3}). We also point out the breakdown of the GUT relation of the gaugino masses, which will have important consequence for the phenomenological analysis in section \ref{sec:4}. In section \ref{sec:5} we discuss the collider signals of the models. The final section (section \ref{sec:6}) is devoted to conclusion and discussion. The Appendix contains the calculation of the mass spectrum in a model of semi-direct gauge mediation.

\section{General Arguments}\label{sec:2}

In this section we collect various arguments which demonstrate the difficulty of generating large enough gaugino masses in a large class of non-minimal GMSB models~\footnote{See also Ref.~\cite{Kitano:2010fa} for a recent review.}, and also discuss the violation of the GUT relation of the gaugino masses. Most of the discussions in this section are review of known facts, but we hope to give a unified treatment of results scattered in the literature, 
and give some new insights at some points.

\subsection{Smallness of Gaugino Masses}\label{sec:2.1}
In the minimal gauge mediation~\cite{Giudice:1998bp,mGMSB}, sfermion and gaugino masses are of the same order.
If a messenger mass is $M$ and a SUSY breaking field $X$ has an $F$-term $X=\h^2F$ with $|F| \ll |M|^2$, the sfermion and gaugino masses are  
generated by the low energy operators~\cite{Giudice:1997ni},
\beq
\d{\cal L}_{\rm sfermion} &\sim& - \int \! d^4 \h \,  \frac{X^\dagger X}{|M|^2} \f^\dagger \f , \label{eq:kahlersferm}  \\
\d{\cal L}_{\rm gaugino}  &\sim&  \int\! d^2 \h\,\frac{X}{M} W^\a W_\a ,
\eeq
where $\f$ is a minimal SSM (MSSM) chiral field and $W_\a$ is a field strength of the MSSM gauge multiplet~\footnote{
In this paper we omit gauge factors $e^{-2V}$ in the matter kinetic terms for simplicity.}. Here we neglected gauge coupling factors.
Both of the above operators lead to the MSSM soft masses of order $F/M$. 
 
We now explain that the statement above is not true in more generic gauge mediation models. As we will see, the reasonings for this are basically holomorphy, R-symmetry and the vacuum structure \footnote{Most of these arguments do not directly apply to $D$-term supersymmetry breaking. However, in practice gauginos are still suppressed by powers of $\sqrt{D}/M$, as we will see later in section~\ref{subsec:D-term}.}.

The first argument uses holomorphy.
In generic gauge mediation models, the sfermion masses are generated by K\"ahler potential operators similar to Eq.~(\ref{eq:kahlersferm}).
Because the K\"ahler potential is not protected from quantum corrections,
and the operator Eq.~(\ref{eq:kahlersferm}) is invariant under any internal symmetry, the operator almost always exists.  
Then the sfermion masses are generated at the leading order of 
the SUSY breaking scale $F$, i.e. $m_{\rm sfermion} \propto |F|$.

By contrast, the gaugino masses are not always generated at the order $F$.
To generate the gaugino masses, we need a low energy operator of the form
\beq
\int \! d^2\h\, H(X)W^\a W_\a, \label{eq:holoop} 
\eeq
where $H(X)$ is a holomorphic function of the SUSY breaking field $X$ (and possibly some other hidden sector light 
fields, which is not explicitly shown here). The gaugino masses are then given by
\beq
M_{\rm gaugino} \sim F \frac{\q H}{\q X}. \label{eq:gauginoformula}
\eeq
Operators of the form Eq.~(\ref{eq:holoop}) are quite severely constrained by holomorphy and symmetry~\cite{Seiberg:1993vc}.
For example, if $X$ is charged under some hidden sector symmetry and if there are no other light fields, we cannot 
have a holomorphic function $H(X)$ other than a constant function. Even if there is some function respecting symmetry, 
often it can be shown by taking weak coupling limits that such a function cannot appear~\cite{Seiberg:1993vc} in Eq.~(\ref{eq:holoop}).

Strictly speaking, there is in fact a contribution to the gaugino masses  
at the leading order of $F$, which is not related to the operator (\ref{eq:holoop}).
Nearly massless fields charged under the MSSM gauge fields can generate a non-local operator of the form~\cite{Shifman:1986zi}
\beq
\int \! d^4\h \, R(X,X^\dagger) W^\a \frac{D^2}{\q^2} W_\a , \label{eq:nonholoop}
\eeq
where $D$ is a superspace derivative, and $R$ is a function of $X$ and $X^\dagger$ as well as momentum and coupling constants. 
If $R$ is holomorphic, we can rewrite Eq.~(\ref{eq:nonholoop}) in the form Eq.~(\ref{eq:holoop}) by using $\bar{D}^2D^2W_\a=16\q^2W_\a$.
But now $R$ is not protected by holomorphy, so we can obtain the gaugino masses at the leading order of $F$ from the operator Eq.~(\ref{eq:nonholoop}).
Unfortunately, this contribution to the gaugino masses is suppressed by higher loops in the MSSM gauge couplings. 
This is because nearly massless fields present in the low-energy theory are the MSSM fields only, and hence additional loops involving those fields are necessary 
to generate the gaugino masses.  
In Ref.~\cite{ArkaniHamed:1998kj}
it was shown that such a contribution is generated only at 3-loop order in the MSSM gauge couplings, regardless of the dynamics in the hidden sector.
Thus this contribution is always small compared with the sfermion masses. This is called the ``gaugino mass screening'' in Ref.~\cite{ArkaniHamed:1998kj}.
For example, in the case of the minimal gauge mediation, the ratio of the 1-loop contribution and the 3-loop contribution to the gaugino masses $M_a$ 
($a=1,2,3$) is given by
\beq
\frac{M^{\rm 3-loop}_{a}}{M^{\rm 1-loop}_{a}} = \sum_{i,b} 8t_a(i)C_b(i)\int^{M_{\rm mess}}_{m_{\rm soft}} \frac{d\mu}{\mu}\left(\frac{g_b^2(\mu)}{16\pi^2}\right)^2,  \label{eq:onevsthreeloop}
\eeq
where $g_a$ ($a=1,2,3$) are the gauge coupling constants of $U(1)_Y,SU(2)_L,SU(3)_C$ respectively, and
$M_{\rm mess}$ and $m_{\rm soft}$ are the messenger and the soft mass scale respectively, 
$t_a(i)$ and $C_a(i)$ are the Dynkin index and quadratic Casimir invariant of the MSSM chiral field $\f_i$, and $i$
runs over the MSSM field species. 
In fact, for a large messenger mass scale $M_{\rm mess}$, the ratio (\ref{eq:onevsthreeloop}) for the gluino becomes as large as
${\cal O}(0.1)$. The 3-loop contribution is model-dependent and is often much smaller. 
See Eqs.~(\ref{eq:threeloopinsemi}, \ref{eq:Aterminsemi}) for another example.
However, the 3-loop contribution may prevent too large splittings in the split GMSB.

Another argument comes from R-symmetry. It was shown in Ref.~\cite{Nelson:1993nf} that any generic, calculable SUSY breaking model must have an R-symmetry, which in turn implies that the Majorana gaugino masses vanish if the R-symmetry is not broken \footnote{This is not the case if the gauginos have Dirac gaugino masses.}. It is therefore necessary to break R-symmetry. For example, there is a class of direct mediation models which break R-symmetry spontaneously.  However, it was noticed in the study of these models that the gaugino masses in Eq.~(\ref{eq:gauginoformula}) vanish  to leading order, {\it even if R-symmetry is spontaneously broken}.

The basic mechanism underlying this phenomena, based on the vacuum structure of the theory, is clarified in Ref.~\cite{Komargodski:2009jf}.  
Consider a SUSY breaking model which can be described by a weakly coupled O'Raifeartaigh-type SUSY breaking model at low energies.
We gauge a subgroup of the flavor symmetry of the model to construct a direct gauge mediation model.
The relevant part of the low energy superpotential is given by
\beq
W=-F^\dagger X +\frac{1}{2}(M_{ab}+\l_{ab}X)\P_a \P_b +\cdots ,
\eeq
where $\P_a$ are messenger fields in the real representation of the MSSM gauge groups (e.g. $\bf 5+\bar{5}$ of $SU(5)_{\rm GUT}$). 
$X$ is a SUSY breaking field, and the lowest component of $X$ spans a flat direction 
if the K\"ahler potential and the superpotential contain only renormalizable terms~\cite{Ray:2006wk}.
In this model, the function $H(X)$ in Eq.~(\ref{eq:holoop}) is given by
\beq
H(X) \sim \log\det(M+\l X).
\eeq

In Ref.~\cite{Komargodski:2009jf}, it was proved that if the vacua spanned by $X$ are true minima of the low energy effective theory~\footnote{Note that 
the vacua need not be stable in the full theory including non-perturbative effects. 
Only the stability of the vacua in the low energy theory at the tree level is required in the following discussion.},
then $\det(M+\l X)=\det(M)$, implying Eq.~(\ref{eq:gauginoformula}) to vanish.
Roughly, a sketch of their proof is given as follows. Suppose that $\det(M+\l X)$ depends on $X$. Then, since $\det(M+\l X)$ is a polynomial in $X$, 
it has a zero at some point $X_0$, $\det(M+\l X_0)=0$. At the point $X_0$, SUSY invariant mass matrix of the messengers $M+\l X_0$
has a zero eigenvector and a SUSY invariant mass vanishes in this direction. However, there is SUSY breaking mass of the messengers from the 
$F$-term of $X$, which make the messengers tachyonic in the zero eigenvector direction. This implies the existence of lower lying vacua in the tachyonic direction, contradicting to the initial assumption that the space spanned by $X$ is the lowest energy state in the low energy theory.
Thus, we can conclude that $\det(M+\l X)$ should not depend on $X$. Note that the vanishing of Eq.~(\ref{eq:gauginoformula})
is established not by symmetry, but by the vacuum structure of the theory. Thus the above argument strengthens the difficulty of generating the gaugino masses.

The above argument could be further generalized (far less rigorously) to a wider class of (semi-)direct gauge mediation models.
Suppose that the holomorphic function $H(X)$ in Eq.~(\ref{eq:holoop}) depends on the SUSY breaking field $X$. The SUSY breaking field $X$ 
may be taken as a composite operator (e.g. as in the 3-2 model of SUSY breaking).
If $H(X)$ is not put by hand and is generated by (perturbative or non-perturbative) dynamics of the model, 
the typical behavior of $H(X)$ at $X \to \infty$ is that $H(X) \sim \log X$.
This is because $H(X)$ can be seen as a field-dependent gauge coupling constant and the dependence of the gauge coupling constant on a mass scale 
is logarithmic at high energies. In this case, we have $H(X)/X \to 0$ at $X \to \infty$, and such a holomorphic function $H(X)$ must have a singularity
at some point $X_0$. In quantum field theory, it is often the case that the singularity in the gauge kinetic function is related to the appearance of 
massless charged particles at the singular point, if the effective theory is well defined near that point~\footnote{
$X$ is not necessarily a moduli field, and hence the effective field theory analysis may break down completely near  the singular point in some cases. 
One of the most notable examples is given by the original models of direct gauge mediation~\cite{directGMSB}. There, a runaway superpotential is dynamically
generated in the SUSY breaking sector, and the runaway potential becomes singular precisely at the point $X_0$ at which $H(X)$ becomes singular. 
Thus these models circumvent our argument.
}. 
Let us call the charged particles as ``messengers'', and denote them by $\P$.
Near the point $X \sim X_0$, the effective superpotential of $X$ and $\P$ will be schematically of the form,
\beq
W \sim (X-X_0)g(X)\Psi\Psi -f(X),
\label{eq:effsuper}
\eeq
where the first term ensures the masslessness of $\P$ at the point $X_0$, and the second term (with the derivative $f'(X)\ne 0$) represents the SUSY breaking.
The superpotential Eq.~(\ref{eq:effsuper}) suggests that the SUSY could be restored (though it cannot necessarily be justified by general reasoning) 
by ``messenger condensation'' ,
\beq
\vev{\P\P} \sim f'/g,
\eeq
at $X \sim X_0$. If the SUSY is restored completely by the messenger condensation, 
the above argument applies to any nonsingular K\"ahler potential of $X$, and hence $X$ is not
necessarily a pseudo moduli. If the SUSY is not completely restored, 
the precise position of the vacuum of the model depends on the K\"ahler potential. 
Although the above argument is not a proof of a no-go theorem, we believe that 
it shows a general difficulty in generating the leading term in (semi-)direct gauge mediation where 
the SUSY breaking vacuum is (sufficiently) stable.


\bigskip

In any model, there is no difficulty in generating the gaugino masses at the higher orders in $F$ if $R$-symmetry is broken in the hidden sector. 
For example, we can write an operator of the form
\beq
\int \! d^4\h\, \frac{X^\dagger X X^\dagger D^2X}{|M|^6}W^\a W_\a,\label{eq:higher}
\eeq
where we assumed $|X| \ll |M|$ and $|F| \ll |M|^2$, with $X$ the Vacuum Expectation Value (VEV) of the superfield $X$ (We often use 
the same symbol for a chiral superfield and its lowest component). 
This operator is not protected by holomorphy or symmetry,
and generates the gaugino masses of order
\beq
M_{\rm gaugino} \sim \frac{|F|^2 F X^\dagger}{|M|^6}. 
\eeq
One may think that we can achieve the gaugino masses comparable to the sfermion masses if we increase $X$ and $F$
such that $|X| \sim |M|$ and $|F| \sim |M|^2$. However, typically there are numerical suppressions in operators such as Eq.~(\ref{eq:higher}),
and in some cases there are additional hidden sector loop suppressions, as we will see in examples in the next section.  
Thus, the gaugino masses are suppressed even if we take $|X| \sim |M|$ and $|F| \sim |M|^2$. 

In some models \cite{Kitano:2006xg,Ibe:2009bh} we can deform the original theory, e.g. by introducing new terms in the superpotential, to generate the gaugino masses larger than those in the original
theory. In these examples the gaugino masses depend on the deformation parameters, and the gaugino and the sfermion masses are independent parameters.
Typically, the gaugino masses are smaller than the sfermion masses, and become comparable when the theory resembles a minimal gauge mediation.

Finally, let us comment on the $\mu$ term of the MSSM. In this paper we do not discuss the origin of the $\mu$ term (and $B\mu$ term), 
but for a correct electroweak symmetry breaking to occur, the $\mu$ term must have a specific value 
which is determined by other parameters. Very roughly, in gauge mediation $\mu$ is of order
\beq
|\mu|^2 \sim -\frac{1}{2}m^2_Z-m^2_{\tilde L}+\frac{12 |y_t|^2}{16\pi^2}m^2_{\tilde Q}\log\left(\frac{M_{\rm mess}}{m_{\tilde Q}}\right)+{\cal O}(1/\tan^2\b),\label{eq:muorder}
\eeq
where $m_{Z}$ is the $Z$ boson mass, $m_{\tilde L}$ is the left-handed slepton mass, $m_{\tilde Q}$ is the squark mass, $y_t$ the top Yukawa coupling,
and $M_{\rm mess}$ the messenger mass scale. Eq.~(\ref{eq:muorder}) shows that generically $\mu$ is of the order of the sfermion masses, unless there is a non-trivial fine-tuning between the parameters.
Therefore, in split GMSB, the Higgsino mass is the same order as the sfermion masses, and the only light superparticles are the three gauginos.

\subsection{Violation of the GUT Relation}\label{sec:GUTtukinuke}
In a minimal gauge mediation model, there is a relation of the gaugino masses known as the GUT relation ($M_{\tilde B}:M_{\tilde W}:M_{\tilde g} \simeq \alpha_1 : \alpha_2 : \alpha_3$).
This relation, however, is violated in gauge mediation models in which the leading $F/M$ term of the gaugino masses vanish.
In this section, we discuss the violation of the GUT relation of the gaugino masses and
show that generically the gluino tends to be lighter, sometimes becoming the NLSP (although the Wino NLSP may also be possible in some cases).
We discuss two reasons for the violation of the GUT relation; one is the contribution of the SM gauge interactions to the renormalization group (RG) evolution
of messenger papameters (see e.g.~Refs.~\cite{Nomura:1997uu,Ibe:2007wp}, and especially Ref.~\cite{Hamaguchi:2008yu} which points out a possibility of
the gluino NLSP). The other is the Higgs-Higgsino loop contribution, which is similar to gauge mediation with $\mu$ and $B\mu$ terms corresponding
to the SUSY invariant and non-invariant masses of messengers in gauge mediation. 

\subsubsection{Standard Model RG Effect on Messengers}
For simplicity, we assume that messenger fields are in the $\bf 5 + \bar{5}$ representation of $SU(5)_{\rm GUT}$, but generalizations to other representations is 
straightforward.
In the case of $\bf 5 + \bar{5}$ messengers, 
there are ``$\bf 3$ messengers'' and ``$\bf 2$ messengers'' under the decomposition ${\bf 5} \to {\bf 3}_{-\frac{1}{3}}+{\bf 2}_{\frac{1}{2}}$.
We denote the masses of ``$\bf 3$ messengers'' and ``$\bf 2$ messengers'' by  $M_d$ and $M_{\ell}$, respectively.
In general, the values of $M_d$ and $M_{\ell}$ are different.
Even if $M_d=M_{\ell}$ at the GUT scale, 
$M_d \neq M_{ \ell}$  at the messenger scale because of RG effects.
Generally, $M_d$ tends to be larger than $M_{\rm \ell}$ because of the strong $SU(3)_C$ interaction.
 
Let us consider the RG evolution of a minimal gauge mediation model~\cite{Dimopoulos:1996ig} as an example. 
The superpotential corresponding to the coupling between SUSY breaking field $X$ and 
the messenger fields $\P_{\ell}, ~\P_{d}$ is given as
\beq
W  = \l_d X \tilde{\P}_{d} \P_d +  M_d \tilde{\P}_{d} \P_d + 
\l_{\ell} X \tilde{\P}_{\ell} \P_{\ell} +  M_{\ell} \tilde{\P}_{\ell} \P_{\ell}.\eeq
In this case, the gaugino masses are given by
 \beq \label{eq:gaugino-mass}
 M_a = \frac{\a_a}{4\pi} \L_{Ga}~~~(a=1,2,3),
 \eeq
where $\L_{Ga}$ are given as
\beq
\L_{G1}&=&\left(\frac{2}{5}\frac{\l_d F}{M_d} + \frac{3}{5}\frac{\l_{\ell} F}{M_{\ell}}\right)+{\cal O}(F^3/M^5),\\
\L_{G2}&=&\frac{\l_{\ell} F}{M_{\ell}}+{\cal O}(F^3/M^5),\\
\L_{G3}&=&\frac{\l_{d} F}{M_{d}}+{\cal O}(F^3/M^5).
\eeq
The masses $M_d,M_{\ell}$ and the Yukawa couplings $\l_d,\l_{\ell}$ obey the RG equations  given by~\footnote{For the convention of anomalous dimensions,
we follow Ref.~\cite{Martin:1997ns}.}
\beq
\frac{\partial \log M_\chi}{\partial \log \mu} &=& \g_{\P_\chi}+\g_{\tilde{\P}_\chi}, \\
\frac{\partial \log \l_\chi}{\partial \log \mu} &=& \g_{\P_\chi}+\g_{\tilde{\P}_\chi}+\g_X,~~~~~(\chi=d,\ell)
\eeq
where $\g_{\P_\chi},\g_{\tilde{\P}_\chi}~(\chi=d,\ell),$ and $\g_X$ are the anomalous dimensions of $\P_\chi,\tilde{\P}_\chi$ and $X$, respectively.
As one can see from the RG equations,
the ratio of the messenger Yukawa coupling and mass, $\l_{\chi}/M_{\chi}$, is the same for $\P_d$ and $\P_\ell$ at any energy scale because of the cancellation
of the contribution of the messenger anomalous dimensions.
Thus, once $\l_{d}/M_{d} = \l_{\ell}/M_{\ell}$ is imposed at the GUT scale, the relation
$\Lambda_1 = \Lambda_2 = \Lambda_3$ is kept at the messenger scale and the GUT relation is maintained, if the leading term of order $\l F/M$ is dominant.

The above cancellation of the messenger anomalous dimension contribution is related to the fact that the leading term of order ${\cal O}(F/M)$ is generated by a holomorphic operator Eq.~(\ref{eq:holoop}). The RG evolution of the parameters in the messenger sector comes from the wave function renormalization
of the fields in the K\"ahler potential. However, the holomorphic operator Eq.~(\ref{eq:holoop}) can be determined by holomorphic quantities only, 
without reference to the wave function renormalization, due to the non-renormalization theorem. Thus we can neglect any perturbative corrections in 
Eq.~(\ref{eq:holoop}), and hence we can consider as if Eq.~(\ref{eq:holoop}) is generated at the GUT scale with the messengers integrated out.
(But note that the VEV of the $F$-term of $X$ depends on the wave function renormalization, because we have to solve the equation of motion to get the $F$-term.) Thus, generically the GUT relation is maintained in gauge mediation models where the leading order gaugino masses do not vanish, if the GUT relation
is maintained at the GUT scale.

However, the above cancellation does not hold for the higher order terms in the gaugino masses, which are generated by K\"ahler potential
operators like Eq.~(\ref{eq:higher}). Consider the term of order
\beq
\frac{\l F}{M}
\left(
\frac{\l F}{M^2}
\right)^{p}. \label{eq:p-term}
\eeq
The ratio of the term~($\ref{eq:p-term}$) for $\P_d$ to the term for $\P_\ell$ at the messenger mass scale is given by
\beq
\left[\frac{\l_d F}{M_d}\left(\frac{\l_d F}{M_d^2}\right)^{p}\right]\left[\frac{\l_d F}{M_d}\left(\frac{\l_d F}{M_d^2}\right)^{p}\right]^{-1}=
\exp \left[p\int^{M_{\rm GUT}}_{M_{\rm mess}} \frac{d\mu}{\mu} (\g_{\P_d}+\g_{\tilde{\P}_d}- \g_{\P_\ell}-\g_{\tilde{\P}_\ell})\right],
\eeq
where $M_{\rm GUT} \sim 10^{16}~$GeV is the GUT scale, $M_{\rm mess}$ is the messenger mass scale, $M_{\rm mess} \sim M_d \sim M_\ell$.
We have assumed that $M_d = M_\ell$ and $\l_d=\l_\ell$ at the GUT scale, and ignored the small logarithm corrections arising from the mismatch of 
$M_d$ and $M_l$.

In the case of the minimal gauge mediation, the violation of the GUT relation in the higher order terms is not so important because the leading term $\l F/M$
is dominant for most of the parameter space of the model. However, in gauge mediation models where the leading term in the SUSY breaking scale $F$
vanishes, only higher order terms in $F$ are present and the GUT relation of the gaugino masses is violated by a non-negligible amount.
In many (though not all) models with $\bf 5 + \bar{5}$ messengers, it is possible to parametrize the gaugino masses as  
\beq
M_1&=&\Lambda\frac{\alpha_1}{4\pi}\left(\frac{2}{5}r^{-p} + \frac{3}{5}\right),\\
M_2&=&\Lambda\frac{\alpha_2}{4\pi},\\
M_3&=&\Lambda\frac{\alpha_3}{4\pi}r^{-p},
\eeq
where $p$ is some number, and $r$ is given by
\beq
r=\exp \left[-\int^{M_{\rm GUT}}_{M_{\rm mess}} \frac{d\mu}{\mu} (\g_{\P_d}+\g_{\tilde{\P}_d}- \g_{\P_\ell}-\g_{\tilde{\P}_\ell})\right].\label{eq:rdef}
\eeq
The GUT relation is retained if $p=0$ or $r=1$, but that is not true for non-minimal models. We will see explicit examples in the next section.
We will also discuss an example where the above parametrization is not valid and the Wino may become the NLSP (see section~\ref{sec:stableftermgmsb}).

Let us calculate $r$ at 1-loop level~\cite{Dimopoulos:1996ig}.
We assume that the only source of the violation of the GUT relation is the SM gauge interaction, and we neglect other interactions (for example, Yukawa interactions in the hidden sector). 
Then, we obtain
\beq
(\g_{\P_d}+\g_{\tilde{\P}_d}) - (\g_{\P_\ell}+\g_{\tilde{\P}_\ell})=\left(-\frac{8}{3}\frac{\a_3}{2\pi}-\frac{2}{15}\frac{\a_1}{2\pi} \right)-
\left(-\frac{3}{2}\frac{\a_2}{2\pi}-\frac{3}{10}\frac{\a_1}{2\pi} \right).\label{eq:messanomconb}
\eeq 
The 1-loop RG equations for the gauge couplings above the messenger mass scale are given by
\beq
\frac{\q}{\q \log \mu} \left( \frac{2\pi}{\a_a} \right)= -b_a,\label{eq:1loopgaugeRG}
\eeq
where $b_1 = 33/5+N_5$, $b_2 = 1+N_5$, and $b_3 = -3+N_5$, with $N_5$ the messenger number. 
Using Eqs.~(\ref{eq:rdef},\ref{eq:messanomconb},\ref{eq:1loopgaugeRG}), we obtain 
\beq
r &=& \left(\frac{\alpha_{\rm GUT}}{\alpha_{3,{\rm mess}}}\right)^{8/(3b_3)}
 \left(\frac{\alpha_{\rm GUT}}{\alpha_{2,{\rm mess}}}\right)^{-3/(2b_2)}
 \left(\frac{\alpha_{\rm GUT}}{\alpha_{1,{\rm mess}}}\right)^{1/(6b_1)} ~~~~~~~~~(N_5 \ne 3), \\
r &=& \left(\frac{M_{\rm GUT}}{M_{\rm mess}}\right)^{4\alpha_{\rm GUT}/(3\pi)}
 \left(\frac{\alpha_{\rm GUT}}{\alpha_{2,{\rm mess}}}\right)^{-3/(2b_2)}
 \left(\frac{\alpha_{\rm GUT}}{\alpha_{1,{\rm mess}}}\right)^{1/(6b_1)} ~~~(N_5 = 3),
\eeq
where $\alpha_{\rm GUT}$ is the gauge coupling at the GUT scale and $\alpha_{a,{\rm mess}}$ are evaluated at the messenger scale.
Approximately, the relevant gauge couplings are given as
\beq
\alpha_{\rm GUT}^{-1} &\sim& 25 - \frac{N_5}{2\pi} \log\left(\frac{M_{\rm GUT}}{M_{\rm mess}}\right)+ \frac{5}{4\pi} \log(R),\\
\alpha_{i,{\rm mess}}^{-1} &=& \alpha_{\rm GUT}^{-1}  + \frac{b_i}{2\pi}  \log\left(\frac{M_{\rm GUT}}{M_{\rm mess}}\right),
\eeq
where $R$ is the ratio of the sfermion mass scale to the gaugino mass scale, and we have assumed that $\L \sim 100~$TeV in the gaugino mass formulae.

\begin{figure}[htbp]
\begin{tabular}{cc}
\begin{minipage}{0.48\hsize}
\begin{center}
\epsfig{file=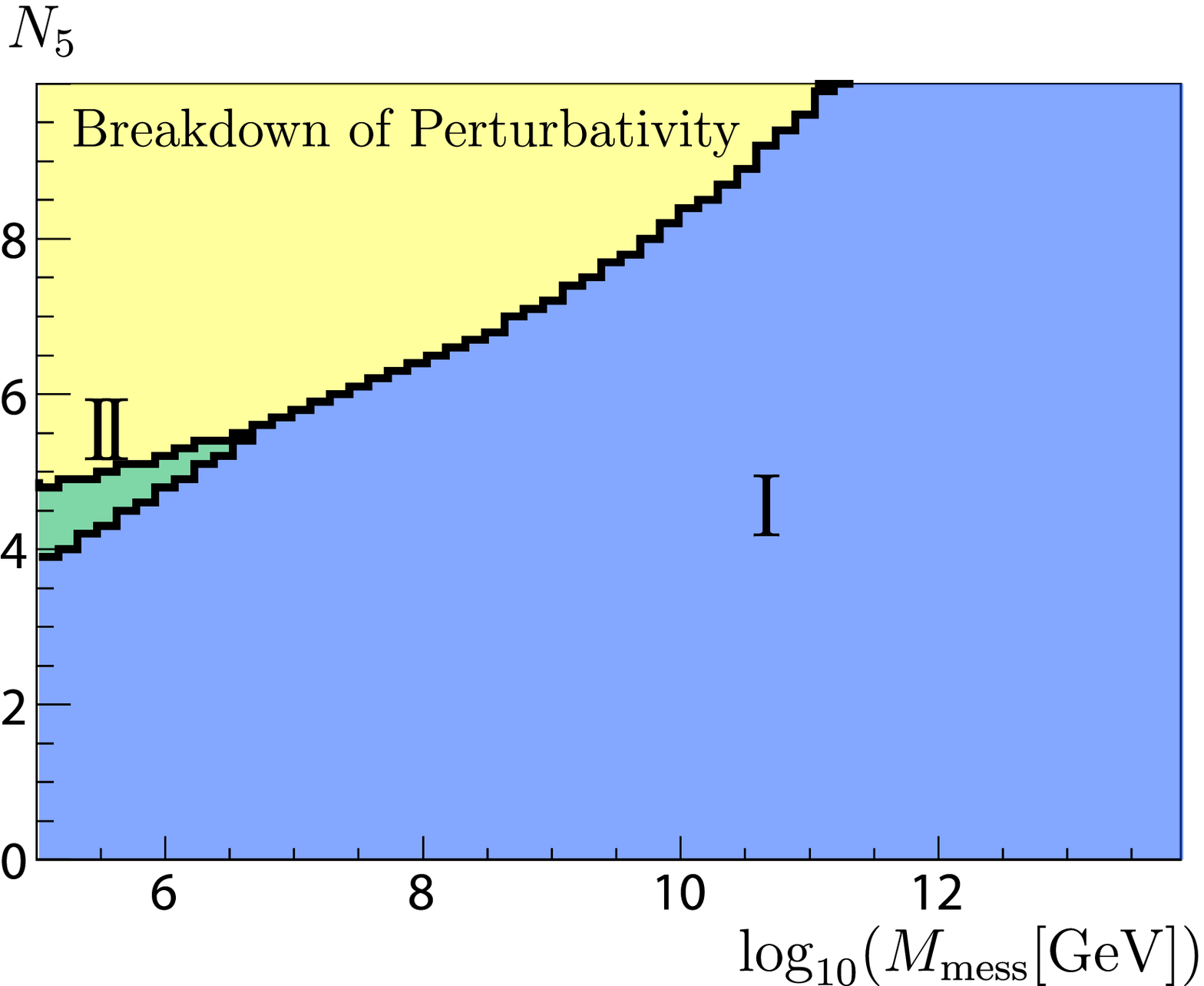 ,scale=.45,clip}
(a) $p=2$
\end{center}
\end{minipage}
\begin{minipage}{0.48\hsize}
\begin{center}
\epsfig{file=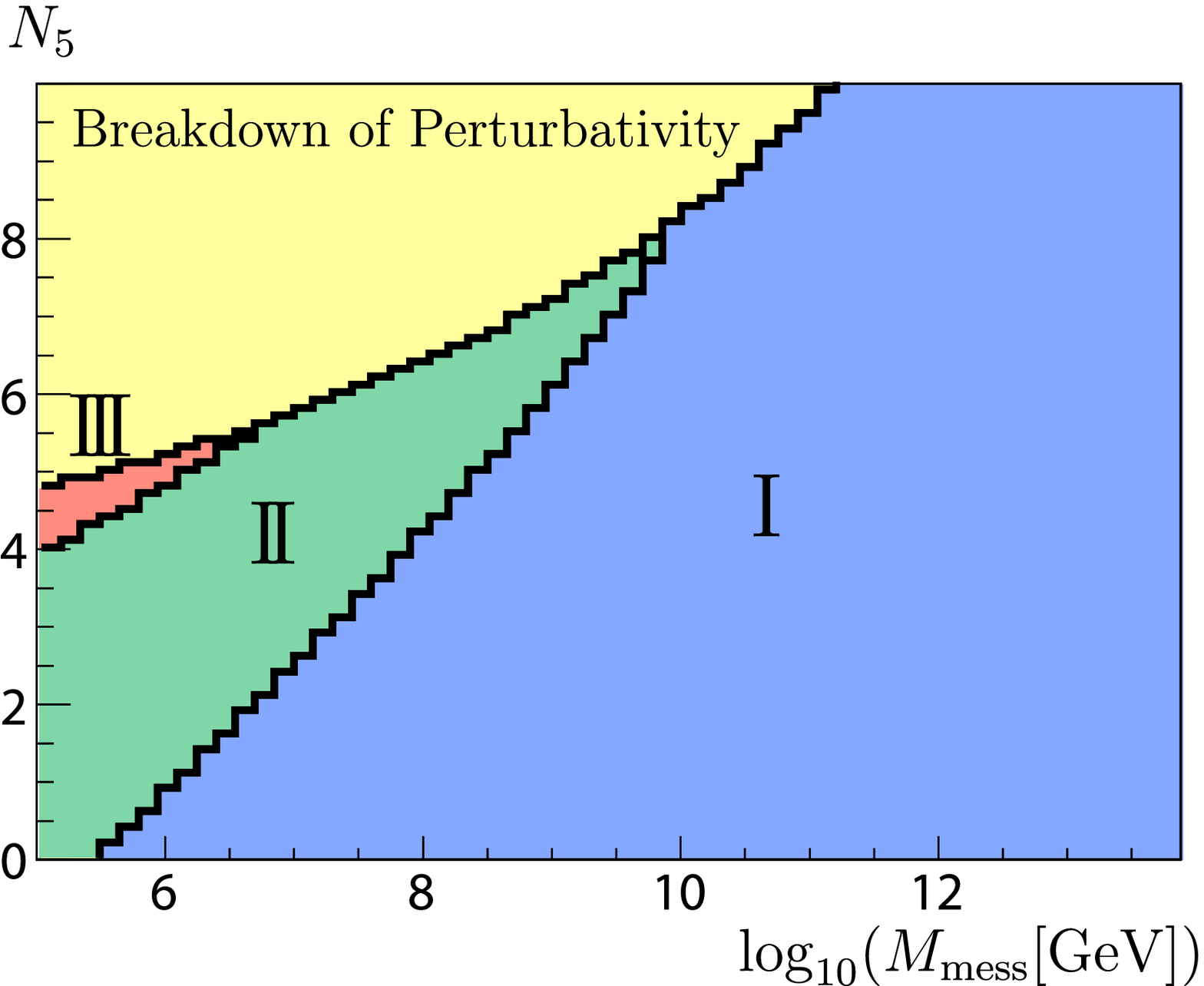 ,scale=.45,clip}
(b)$p=4$
\end{center}
\end{minipage}\\

\\

\begin{minipage}{0.48\hsize}
\begin{center}
\epsfig{file=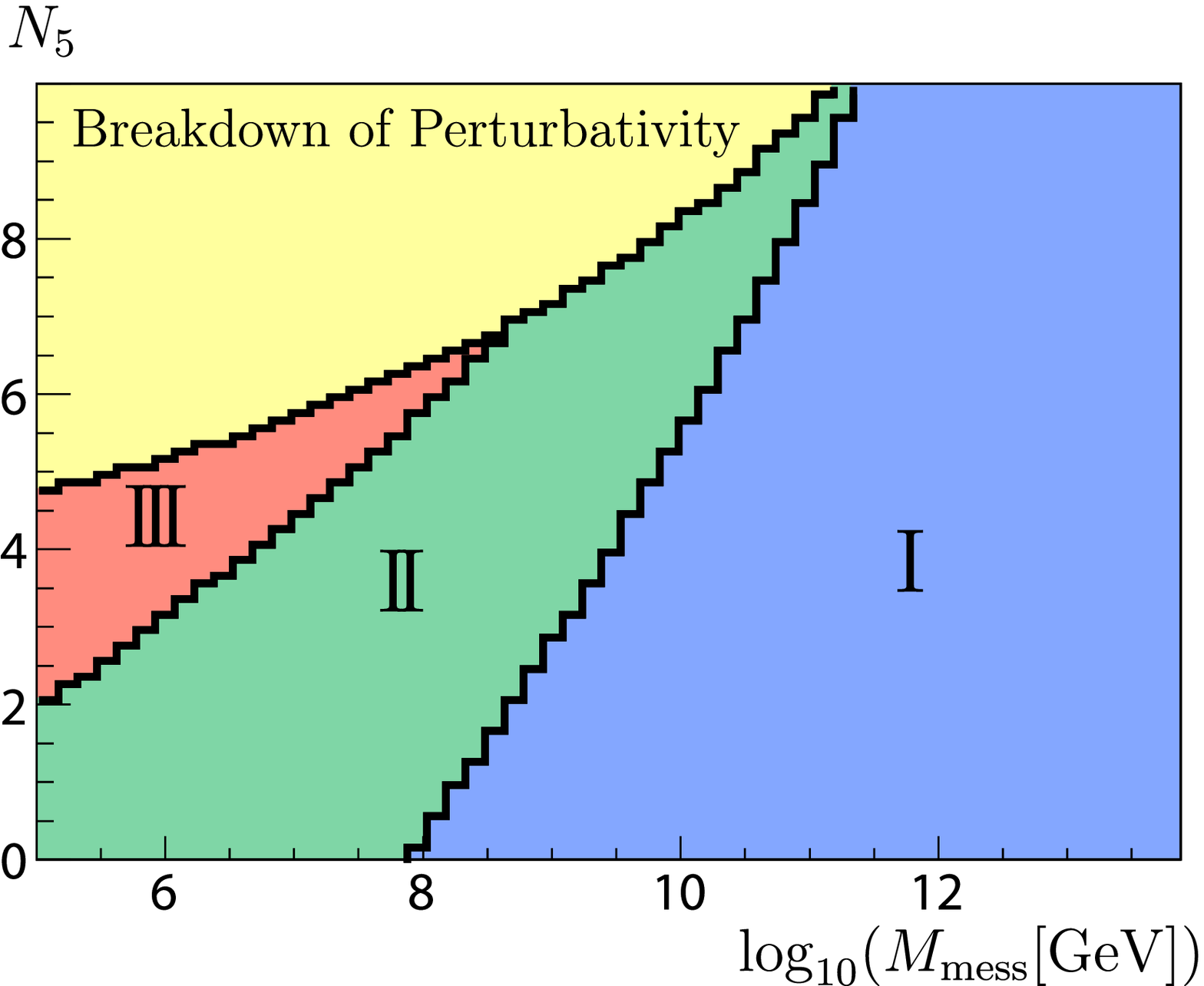 ,scale=.45,clip}
(c) $p=6$
\end{center}
\end{minipage}

\begin{minipage}{0.48\hsize}
\begin{center}
\epsfig{file=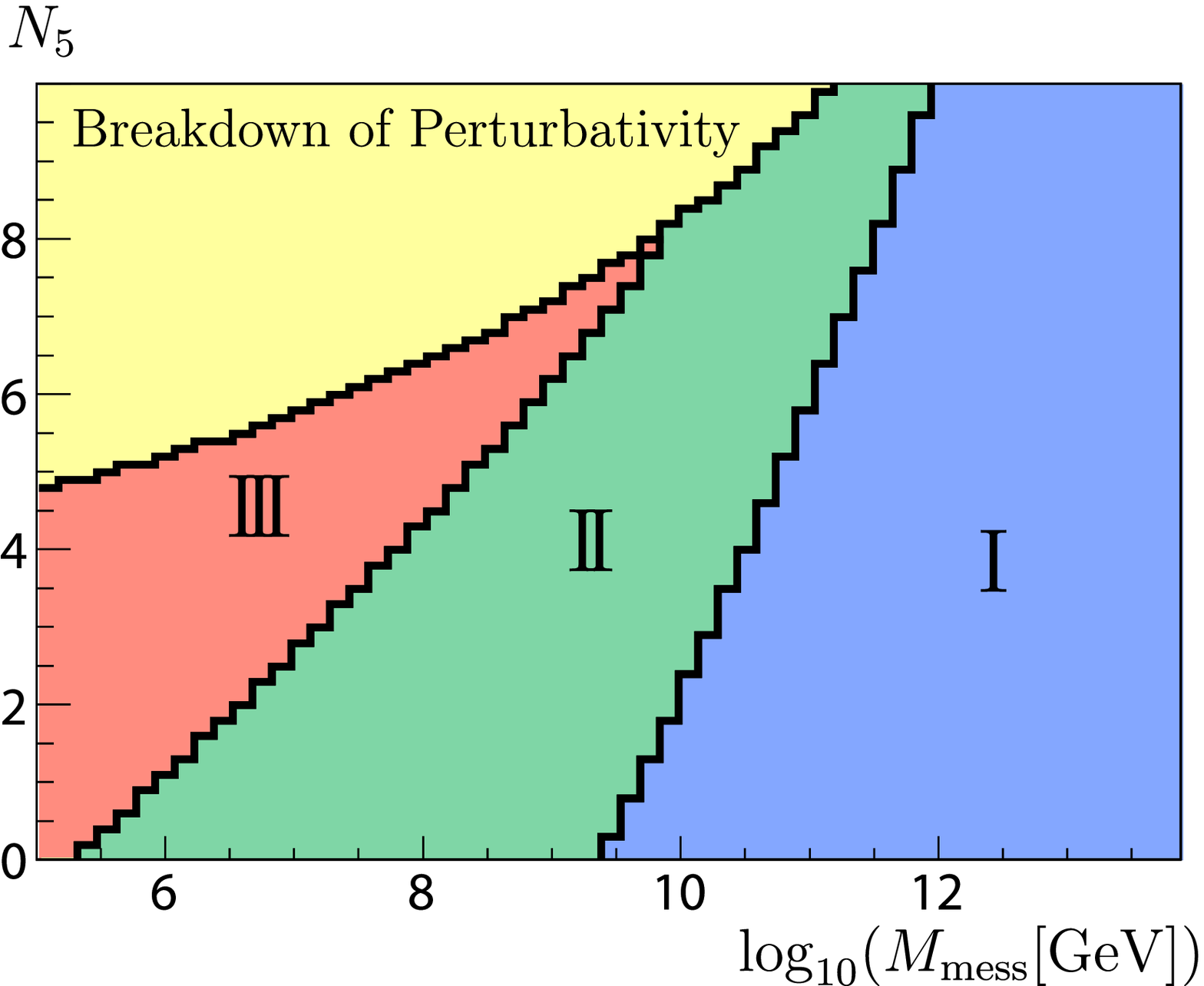 ,scale=.45,clip}
(d) $p=8$
\end{center}
\end{minipage}\\

\\

\end{tabular}
\caption{The relation among the masses of the neutralinos ${\tilde \chi}^0_1\simeq {\tilde B}$, ${\tilde \chi}^0_2\simeq {\tilde W^0}$, and the gluino ${\tilde g}$. 
In region I (blue region), we have $m_{\tilde{\chi}^0_1}<m_{\tilde{\chi}^0_2}<m_{\tilde{g}}$,
region I\!I (green region), $m_{\tilde{\chi}^0_1}<m_{\tilde{g}}<m_{\tilde{\chi}^0_2}$,
and region I\!I\!I (red region), $m_{\tilde{g}}<m_{\tilde{\chi}^0_1}<m_{\tilde{\chi}^0_2}$.
In the yellow region, some of the gauge couplings blow up ($\alpha_i$ is larger than unity) before the GUT unification. }
\label{fig:NLSP}
\end{figure}

In Figs.~\ref{fig:NLSP}, we show the relation between the masses of the gluino and neutralinos. 
The neutralino masses $m_{\tilde{\chi}^0_1}$ and
$m_{\tilde{\chi}^0_2}$ are almost the Bino and Wino mass, respectively, so Figs.~\ref{fig:NLSP} can be seen as a relation among the three gaugino masses.
We use 2-loop beta functions of the gauge couplings and leading order anomalous dimensions of the matter fields.
We take the sfermion masses to be 
\beq
m^2_{\f_i}=\Lambda_S^2 \left\{\left(\frac{\a_1}{4\pi}\right)^2 C_1(i)+\left(\frac{\a_2}{4\pi}\right)^2
C_2(i)+\left(\frac{\a_3}{4\pi}\right)^2C_3(i)\right\}.
\eeq
We set $\Lambda_S=10^6~$TeV, $\L=10^5~$GeV, $M_{\rm mess}=10^6~$GeV and $\tan\beta=10$ as an example.
As for the low energy MSSM spectrum, 
we have used the programs \verb+SOFTSUSY+ 2.0.18~\cite{Allanach:2001kg}.
One can see that the gluino can be light drastically, and in some cases the gluino is lightest among the gauginos.

\subsubsection{Higgs-Higgsino Threshold}
If the scalar particles are much heavier than the gauginos,
the second ``gauge mediation'' from Higgs-Higgsino loops is possibly important.
The corrections from the Higgs-Higgsino loops are written as \cite{Pierce:1996zz}
\beq
\Delta M_1 &=& \frac{3}{5}\frac{\alpha_1}{4\pi} \left( \frac{m_A^2\mu\sin 2\beta }{|\mu^2|-m_A^2} \log\left( \frac{|\mu|^2}{m_A^2} \right) \right),\label{eq:higgs1}\\
\Delta M_2 &=& \frac{\alpha_2}{4\pi} \left( \frac{m_A^2\mu\sin 2\beta }{|\mu^2|-m_A^2} \log\left( \frac{|\mu|^2}{m_A^2} \right) \right),\label{eq:higgs2} \\
\Delta M_3 &=& 0,\label{eq:higgs3}
\eeq
where $m_A$ is the CP odd Higgs scalar mass.
If $R\gg 1$, this correction plays a significant role.
Since $\Delta M_3=0$, the gluino becomes light in the limit $R \gg 1$.

\section{Examples}\label{sec:3}
Let us describe some examples of non-minimal gauge mediation models in order to illustrate the discussions in the previous section in concrete setups.
Readers who are interested only in the final results can proceed directly to the summary at the end of this section.

\subsection{Semi-direct Gauge Mediation in the 3-2 Model}\label{sec:semimi}
Let us consider the semi-direct gauge mediation model proposed in Ref.~\cite{Seiberg:2008qj}.
The model is based on the so-called 3-2 model of SUSY breaking~\cite{Affleck:1984xz}.
There are $SU(3)_{\rm hid} \times SU(2)_{\rm hid}$ gauge groups of the hidden sector, 
and the matter content of the model is listed in Table~\ref{table:1}.
We take a tree-level superpotential
\beq
W_{\rm tree}=h \tilde{d}QL + Ml\tilde{l},
\eeq
where $h$ is a Yukawa coupling, and $M$ is the mass of the fields $l,~\tilde{l}$.

\begin{table}[htbp]
\begin{center}
\caption{The matter content of the semi-direct gauge mediation in 3-2 model. The representations under 
$SU(3)_{\rm hid}\times SU(2)_{\rm hid}\times SU(5)_{\rm GUT}$ are listed.}
\vspace{3mm}
\begin{tabular}{|c|c|c|c|c|c|c|}
\hline 
matter&$Q$&$\tilde{u}$&$\tilde{d}$&$L$&$l$&$\tilde{l}$ \\ \hline
representation&$\bf (3,2,1)$&$\bf(\bar{3},1,1)$&$\bf(\bar{3},1,1)$&$\bf(1,2,1)$&$\bf(1,2,5)$&$\bf(1,2,\bar{5})$ \\ \hline
\end{tabular}
\label{table:1}
\end{center}
\end{table}

$SU(3)_{\rm hid}$ instanton generates a dynamical superpotential~\cite{Affleck:1983mk}, and the effective superpotential 
at low energies is given by
\beq
W_{\rm eff}=\frac{\L_3^7}{\det(Q\tilde{Q})}+h \tilde{d}QL + Ml\tilde{l},
\eeq
where $\L_3$ is the dynamical scale of the $SU(3)_{\rm hid}$ gauge group, and $\tilde{Q}=(\tilde{d},\tilde{u})$.
It is known that SUSY is broken in this effective superpotential, and there are non-vanishing VEVs and $F$-terms of the fields $Q,~\tilde{Q}$ and $L$.
The fields $l$ and $\tilde{l}$ do not play a part in the SUSY breaking, and serve as messenger fields.

In Ref.~\cite{Seiberg:2008qj}, it was shown that the K\"ahler potential of $l$ and $\tilde{l}$ takes the following form
after integrating out the $SU(3)_{\rm hid} \times SU(2)_{\rm hid}$ gauge fields.
The set $(Q,\tilde{Q},L)$ is collectively denoted by $q$, and the generators of $SU(3)_{\rm hid} \times SU(2)_{\rm hid}$ by $T^I$.
Fixing $SU(3)_{\rm hid} \times SU(2)_{\rm hid}$ gauge by unitary gauge conditions $\vev{q}^\dagger T^I \vev{q}=0$ and $\vev{q}^\dagger T^I (q-\vev{q})=0$, 
the effective K\"ahler potential is given by~\footnote{
We almost follow the convention of Ref.~\cite{Wess:1992cp}, with one difference on the normalization of vector multiplets: 
a matter kinetic term is $\F^\dagger e^{-2V}\F$ instead of 
$\F^\dagger e^{V}\F$, and a gauge field strength is $W_\a=-\frac{1}{8}\bar{D}^2(e^{2V}D_\a e^{-2V})$ instead of
$W_\a=-\frac{1}{4}\bar{D}^2(e^{-V}D_\a e^{V})$. A gauge kinetic term is $\int d^2\h\frac{1}{4g^2}W^\a W_\a +{\rm h.c.}$, 
where the sum over gauge group index is implicit. This normalization gives the usual component field Lagrangian.}
\beq
K_{\rm eff} &=& |l|^2+ |\tilde{l}|^2 +K_{\rm 0-loop}+K_{\rm 1-loop}+\cdots,\\
K_{\rm 0-loop} &=& -2\l^{-1}_{IJ} (q^\dagger T^I q)(l^\dagger T^J l + \tilde{l}^\dagger T^J \tilde{l})+\cdots, \label{eq:0loop}\\
K_{\rm 1-loop} &=& \frac{3}{4}\frac{g^2_{SU(2)_{\rm hid}}}{8\pi^2}( |l|^2+ |\tilde{l}|^2)\log\left(\frac{g^2_{SU(2)_{\rm hid}}[|Q|^2+|L|^2]/2}{\mu^2}\right) +\cdots , \label{eq:1loop}
\eeq
where $\l^{-1}_{IJ}$ is the inverse of the matrix $\l^{IJ} \equiv q^\dagger \{T^I,T^J\}q$, $g_{SU(2)_{\rm hid}}$ the gauge coupling constant
of $SU(2)_{\rm hid}$, and $\mu$ a renormalization scale. $M_V^2 \equiv g^2_{SU(2)_{\rm hid}}[|\vev{Q}|^2+|\vev{L}|^2]/2$ is the 
mass of the $SU(2)_{\rm hid}$ gauge multiplet when we switch off the $SU(3)_{\rm hid}$ gauge coupling%
~\footnote{
Eq.~(\ref{eq:1loop}) is valid when the $SU(3)_{\rm hid}$ coupling $g_{SU(3)_{\rm hid}}$ is much smaller than the $SU(2)_{\rm hid}$ coupling 
$g_{ SU(2)_{\rm hid}}$ so that there is only a negligible mixing between the $SU(3)_{\rm hid}$ gauge fields and the $SU(2)_{\rm hid}$ gauge fields
in the mass eigenstates. 
When $g_{SU(3)_{\rm hid}}$ is not small, we should use a more general formula presented in Ref.~\cite{Grisaru:1996ve}.
(Note that the logarithm of the gauge boson mass matrix cannot be expanded by the couplings.)
In this paper we assume $g_{SU(3)_{\rm hid}} \ll g_{SU(2)_{\rm hid}}$ for simplicity, but our discussions below do not change qualitatively even if this condition
is not satisfied. 
}.
From Eq.~(\ref{eq:1loop}), we see that the coupling $g_{SU(2)_{\rm hid}}$ should be evaluated 
at the renormalization scale $\m \simeq M_V$ to avoid large logarithm. 

After the decoupling of the gauge fields, there are only irrelevant interactions between $l,\tilde{l}$ and the other hidden sector fields $q$. 
Thus we neglect hidden sector interactions and simply replace the fields by their VEVs. We then have
\beq
K_{\rm eff}=\Z_+(|l_1|^2+ |\tilde{l}_1|^2)+\Z_-(|l_2|^2+|\tilde{l}_2|^2), \label{eq:highkahler}
\eeq
where the subscripts 1 and 2 in $l$ and $\tilde{l}$ are the $SU(2)_{\rm hid}$ indices. 
$\Z_\pm$ are given by
\beq
\log \Z_\pm=\log Z +\frac{1}{2}(f\h^2+{\rm h.c.})+(\mp D+d)\h^2\bar{\h}^2,\label{eq:fdDinsemi}
\eeq
with~\cite{Seiberg:2008qj}
\beq
Z\simeq 1,~~~D \simeq1.48v^2,~~~f \simeq 0.226h_gv,~~~ d \simeq 0.572h_gv^2,
\eeq
where we defined $v \equiv h^{6/7}\L_3$ and $h_g\equiv g_{SU(2)_{\rm hid}}^2/8\pi^2$. In the above equations, only the leading order terms in 
$h_g=g_{SU(2)_{\rm hid}}^2/8\pi^2$ are retained.

Let us show~\cite{Seiberg:2008qj} that there is no term of the form Eq.~(\ref{eq:holoop}) in the low energy effective Lagrangian after the decoupling of $l$ and $\tilde{l}$.
This is a good example of the discussion in the previous section, although experts may find this obvious.
Let us take one of the MSSM gauge groups $G=SU(3),SU(2),{\rm or}~U(1)$ and consider the gauge kinetic function of $G$,
\beq
\int \! d^2\h  \, \frac{1}{4}H W^\a W_\a.
\eeq
From the gauge invariance, we know that $H$ is a holomorphic function of $Y\equiv \det Q\tilde{Q}$, $X_1\equiv \tilde{d}QL$, $X_2\equiv \tilde{u}QL$ as well as
$h,~M,~\L_2,~\L_3$ and $\L_G$, where $\L_2$ and $\L_G$ are ``dynamical scales'' 
 of $SU(2)_{\rm hid}$ and $G$ above the scale $M$, respectively. At the 1-loop level, we know from the simple matching of the high energy and low energy
 gauge couplings that
 \beq
 H|_{\rm 1-loop} = \frac{1}{8\pi^2}\log\left(\frac{\mu^b}{M^2\L_G^{b-2}}\right), \label{eq:lowholo}
 \eeq
where $b$ is the  coefficient of the 1-loop $\b$ function below the scale $M$. What we want to show is that Eq.~(\ref{eq:lowholo}) is in fact exact.

For this purpose, let us consider a $U(1)$ global symmetry with a charge assignment 
\beq
l,\tilde{l}:+1,~~~M:-2,~~~\L_2^{-1}:+10,~~~\L_G^{b-2}:+4,
\eeq
with the charges of other fields taken to be zero.
This $U(1)$ is a symmetry of the high energy theory, as one can check by considering anomaly. 
From this $U(1)$ symmetry, we know that $H$ should depend only on the combinations
$\tilde{\L}_2^{4}\equiv M^5\L_2^{-1}$ and $\tilde{\L}_G^{b}\equiv M^2\L_G^{b-2}$. The physical meaning of 
$\tilde{\L}_2$ and $\tilde{\L}_G$ is that they are dynamical scales of $SU(2)_{\rm hid}$ and $G$ below the scale $M$, respectively.
Then $H$ has the form
\beq
H=H(Y,X_1,X_2,h,\L_3,\tilde{\L}_2,\tilde{\L}_G). \label{eq:lowhololo}
\eeq
Let us take a limit $M \to \infty$ with $\tilde{\L}_2$, $\tilde{\L}_G$ and other variables fixed.
This corresponds to a limit where the messengers $l,\tilde{l}$ decouple from the theory {\it with the low energy theory fixed}.
In this limit, the hidden sector and the MSSM sector decouple completely, so $H$ in this limit is given by
\beq
H \to  \frac{1}{8\pi^2}\log\left(\frac{\mu^b}{\tilde{\L}_G^b}\right). \label{eq:lowhololholo}
\eeq
However, in taking this limit, we have fixed all the variables in Eq.~(\ref{eq:lowhololo}), so $H$ should be exactly given by 
the right hand side of Eq.~(\ref{eq:lowhololholo}). In particular, $H$ does not depend on the hidden sector fields $X_1,X_2$ and $Y$.
This completes the proof. Obviously, the proof can be straightforwardly applied for any theory 
as long as ``messenger fields'' (i.e. hidden sector fields charged under the MSSM gauge groups) have only a mass term of the form $Ml\tilde{l}$ and no Yukawa couplings in a superpotential.
This establishes the vanishing of the gaugino masses at the leading order of SUSY breaking scale in semi-direct gauge mediation.

Now we consider the explicit mass spectrum of the MSSM fields.
The mass spectrum of the MSSM fields is calculated in Appendix~\ref{app:A}.
For the time being, we neglect the mass difference between the $\bf 3$ messengers and $\bf 2$ messengers under the decomposition 
${\bf 5} \to {\bf 3}_{-\frac{1}{3}}+{\bf 2}_{-\frac{1}{2}}$. 
We also assume that $M_V $ is larger than the SUSY-breaking scale and the messenger mass scale. 
Then, using Eqs.~(\ref{eq:massformula1}, \ref{eq:massformula2}), the soft masses are given by
\beq
M_a &\simeq&  \frac{g_a^2}{16\pi^2} v\left[0.172h_g^2\left(\frac{v}{M}\right)^2+0.165h_g\left(\frac{v}{M}\right)^4\right] ,\label{eq:gauginom}  \\
m_{\f_i}^2 &\simeq& \sum_{a=1,2,3} \left(\frac{g_a^2}{16\pi^2} \right)^2 C_a(i)  v^2\left(  4.57h_gL_V-3.76\left(\frac{v}{M}\right)^6  \right) \label{eq:sfermionm},
\eeq
where $g_a$ ($a$=1,2,3) are the gauge couplings for the MSSM gauge group $U(1)_Y,SU(2)_L$ and $SU(3)_C$, respectively, 
$C_a(i)$ the quadratic Casimir for the MSSM particle
$\f_i$, and $L_V=\log (M_{V}^2/M^2)$.
Note that the gaugino masses are severely suppressed by powers of $h_g$, $v/M$, 
and also by numerical factors compared with the first term in Eq.~(\ref{eq:sfermionm}),
as discussed in the previous section.
Requiring $m_{\f_i}^2>0$, we obtain
\beq
h_gL_V>0.82\left(\frac{v}{M}\right)^6.
\eeq
Then, we obtain (by taking into account $h_g \ll1$ and $L_V \gg1$ which are assumed to derive the mass formulae)
\beq
M_a <\frac{g_a^2}{16\pi^2}v( 0.188h_g^{5/3}L_V^{2/3}).
\eeq
If we assume that the second negative term in Eq.~(\ref{eq:sfermionm}) is negligible, 
we obtain the upper bound on the ratio of e.g. the Bino mass to 
the right handed selectron mass,
\beq
\frac{M_{\tilde B}}{m_{\tilde e_R}}<0.113h_g^{7/6}L_V^{1/6}.
\label{Reg1}
\eeq
Requiring perturbativity up to the GUT scale, we have $h_g \ll (\log(M_{\rm GUT}/M_V))^{-1}$ with $M_{\rm GUT} \sim 10^{16}~{\rm GeV}$, 
since $SU(2)_{\rm hid}$ is asymptotic non-free. Thus, the gaugino masses are necessarily suppressed compared with the sfermion masses.

As we will discuss later, the gravitino mass is quite important for cosmological considerations. 
The vacuum energy (in global SUSY) is given by~\cite{Seiberg:2008qj} $V \simeq 2.417h^{10/7}\L_3^4=2.417h^{-2}v^4$. Then, the gravitino mass is
\beq
m_{3/2}=\frac{\sqrt{V}}{\sqrt{3}M_{Pl}}=60~\EV \left[\frac{M^{\rm max}_{\tilde B}}{100~{\rm GeV} }\right]^2\frac{1}{hh_g^{10/3}L_V^{4/3}},
\eeq
where $M^{\rm max}_{\tilde B}$ is roughly the upper bound on the Bino mass,
\beq
M^{\rm max}_{\tilde B}\equiv \frac{g_1^2}{16\pi^2}v( 0.188h_g^{5/3}L_V^{2/3}).
\eeq
A light gravitino mass $m_{3/2} \lsim 16~\EV$ is favored by cosmology as we will discuss in the next section,
but note that the gravitino mass is larger than $16~\EV$ even if the coupling of the hidden sector is very strong, $h_g\simeq 1$, due to 
numerical suppression of the gaugino masses. This will have an important consequence for phenomenological considerations in later sections.

Now consider the violation of the GUT relation of the gaugino masses discussed in section~\ref{sec:GUTtukinuke}.
$f,d,$ and $D$ in Eq.~(\ref{eq:fdDinsemi}) are not renormalized by the SM gauge interaction at 1-loop level, and the renormalization of the messenger mass
is the source of the violation of the GUT relation at 1-loop. From Eq.~(\ref{eq:gauginom}), we can see that $p=2$ if the first term of Eq.~(\ref{eq:gauginom})
dominates and $p=4$ if the second term dominates, in the parametrization of section~\ref{sec:GUTtukinuke}. In the above analysis we consider the model
with the messenger number $N_5=2$, and in that case the GUT violation by the RG effect is not so large as to realize the gluino NLSP.
But it is possible to increase the messenger number in the model. Perhaps more importantly, the mass splitting between the gaugino and sfermion masses
is quite large in this model, so the Higgs-Higgsino contribution shown in Eqs.~(\ref{eq:higgs1}-\ref{eq:higgs3}) can become important.

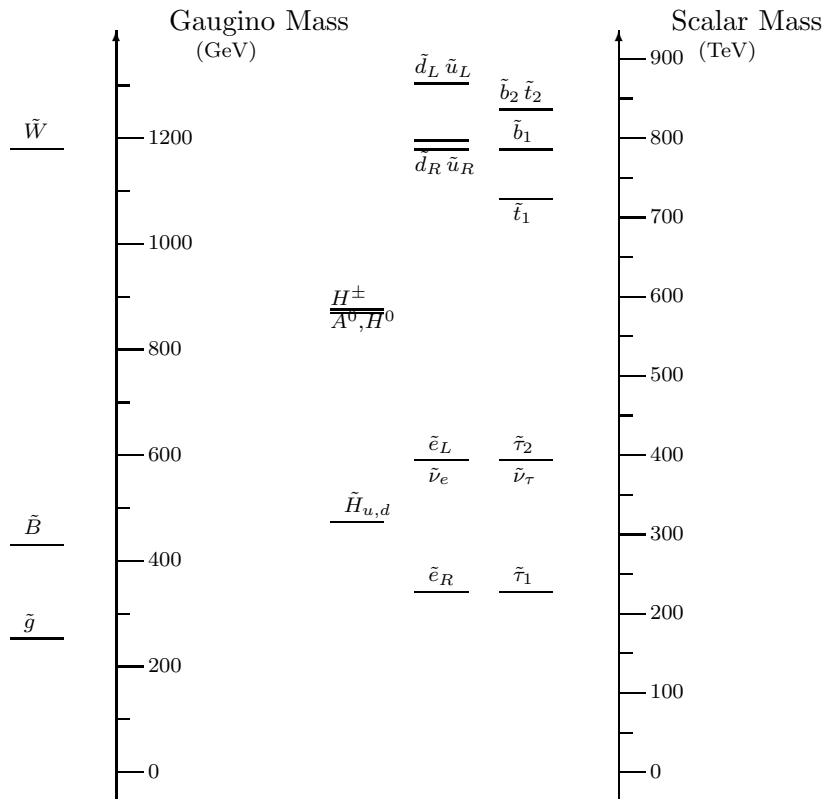
\begin{figure}[htbp]
\begin{center}
\scalebox{1}{
\begin{picture}(280,300.857)(0,0)
\put(250,280.857){\small Scalar Mass }
\put(260,270.857){$\scriptstyle {\rm (TeV)}$ }
\put(230,-10){\vector(0,1){290.857}}
\put(230,0){\line(1,0){10}}
\put(242,-2){$\scriptstyle 0$}
\put(230,30){\line(1,0){10}}
\put(242,28){$\scriptstyle 100$}
\put(230,60){\line(1,0){10}}
\put(242,58){$\scriptstyle 200$}
\put(230,90){\line(1,0){10}}
\put(242,88){$\scriptstyle 300$}
\put(230,120){\line(1,0){10}}
\put(242,118){$\scriptstyle 400$}
\put(230,150){\line(1,0){10}}
\put(242,148){$\scriptstyle 500$}
\put(230,180){\line(1,0){10}}
\put(242,178){$\scriptstyle 600$}
\put(230,210){\line(1,0){10}}
\put(242,208){$\scriptstyle 700$}
\put(230,240){\line(1,0){10}}
\put(242,238){$\scriptstyle 800$}
\put(230,270){\line(1,0){10}}
\put(242,268){$\scriptstyle 900$}
\put(230,0){\line(1,0){5}}
\put(230,15){\line(1,0){5}}
\put(230,30){\line(1,0){5}}
\put(230,45){\line(1,0){5}}
\put(230,60){\line(1,0){5}}
\put(230,75){\line(1,0){5}}
\put(230,90){\line(1,0){5}}
\put(230,105){\line(1,0){5}}
\put(230,120){\line(1,0){5}}
\put(230,135){\line(1,0){5}}
\put(230,150){\line(1,0){5}}
\put(230,165){\line(1,0){5}}
\put(230,180){\line(1,0){5}}
\put(230,195){\line(1,0){5}}
\put(230,210){\line(1,0){5}}
\put(230,225){\line(1,0){5}}
\put(230,240){\line(1,0){5}}
\put(230,255){\line(1,0){5}}
\put(121,175.113){\line(1,0){20}}
\put(121,173.968){\line(1,0){20}}
\put(121,173.968){\line(1,0){20}}
 \put(121,176.968){$\scriptstyle H^{\pm}$}
 \put(121,168.113){$\scriptstyle A^0,H^0$}
\put(0,86.0875){\line(1,0){20}}
\put(5.,90.0875){$\scriptstyle {\tilde{B}}$}
\put(0,236.015){\line(1,0){20}}
\put(5.,240.015){$\scriptstyle {\tilde{W}}$}
\put(0,50.5917){\line(1,0){20}}
\put(5,54.5917){$\scriptstyle {\tilde g}$}
\put(60,280.857){\small Gaugino Mass }
\put(70,270.857){$\scriptstyle {\rm (GeV)}$ }
\put(40,-10){\vector(0,1){290.857}}
\put(40,0){\line(1,0){10}}
\put(52,-2){$\scriptstyle 0$}
\put(40,40){\line(1,0){10}}
\put(52,38){$\scriptstyle 200$}
\put(40,80){\line(1,0){10}}
\put(52,78){$\scriptstyle 400$}
\put(40,120){\line(1,0){10}}
\put(52,118){$\scriptstyle 600$}
\put(40,160){\line(1,0){10}}
\put(52,158){$\scriptstyle 800$}
\put(40,200){\line(1,0){10}}
\put(52,198){$\scriptstyle 1000$}
\put(40,240){\line(1,0){10}}
\put(52,238){$\scriptstyle 1200$}
\put(40,0){\line(1,0){5}}
\put(40,20){\line(1,0){5}}
\put(40,40){\line(1,0){5}}
\put(40,60){\line(1,0){5}}
\put(40,80){\line(1,0){5}}
\put(40,100){\line(1,0){5}}
\put(40,120){\line(1,0){5}}
\put(40,140){\line(1,0){5}}
\put(40,160){\line(1,0){5}}
\put(40,180){\line(1,0){5}}
\put(40,200){\line(1,0){5}}
\put(40,220){\line(1,0){5}}
\put(40,240){\line(1,0){5}}
\put(40,260){\line(1,0){5}}
\put(121,94.691){\line(1,0){20}}
\put(126.,98.691){$\scriptstyle {\tilde{H}_{u,d}}$}
\put(153,260.646){\line(1,0){20}}
\put(153,260.646){\line(1,0){20}}
\put(153,264.646){$\scriptstyle \tilde d_L\,\tilde u_L$}
\put(153,235.717){\line(1,0){20}}
\put(153,239.077){\line(1,0){20}}
\put(153,227.717){$\scriptstyle \tilde d_R\,\tilde u_R$}
\put(153,118.03){\line(1,0){20}}
\put(153,118.03){\line(1,0){20}}
\put(158,110.03){$\scriptstyle \tilde \nu_e$}
\put(158,122.03){$\scriptstyle \tilde e_L$}
\put(153,68.2967){\line(1,0){20}}
\put(158,72.2967){$\scriptstyle \tilde e_R $}
\put(185,216.917){\line(1,0){20}}
\put(190,208.917){$\scriptstyle \tilde t_1$}
\put(185,250.857){\line(1,0){20}}
\put(185,235.696){\line(1,0){20}}
\put(190,239.696){$\scriptstyle \tilde b_1$}
\put(185,250.857){\line(1,0){20}}
\put(185,254.857){$\scriptstyle \tilde b_2\,\tilde t_2$}
\put(185,68.3299){\line(1,0){20}}
\put(190,72.3299){$\scriptstyle \tilde \tau_1$}
\put(185,118.056){\line(1,0){20}}
\put(190,122.056){$\scriptstyle \tilde \tau_2$}
\put(185,118.056){\line(1,0){20}}
\put(190,110.056){$\scriptstyle \tilde \nu_{\tau}$}
\end{picture} }
\end{center}
\caption{Low-energy MSSM spectrum in semi-direct GMSB.
We set $h=0.1$, $g_{SU(2)_{\rm hid}}=1$, $M_\ell=10^9$ GeV, $\Lambda_3 = 3\times 10^9$ GeV and $\tan\beta=5$.
In this parameter choice, the approximations used to obtain the mass formulae Eqs.~(\ref{eq:gauginom},\ref{eq:sfermionm}) are not so good,
and hence the spectrum should be viewed only as representing some qualitative features. 
}
\label{fig:semi-spe}
\end{figure}

In Fig.~\ref{fig:semi-spe}, we show an example of low-energy MSSM spectrum of this model.
The above analysis is done at the messenger scale. However,
in the example, the splitting between the sfermion and gaugino masses is so large that the MSSM quantum corrections (including the 
Higgs-Higgsino loop threshold effect) have sizable impact and the gluino NLSP is realized. 

\subsection{$F$-term Gauge Mediation with Stable Vacuum}\label{sec:stableftermgmsb}
Let us consider the gauge mediation model with the superpotential,
\beq
W=-F^\dagger X+M({\tilde \P}_1\P_2+{\tilde \P}_2\P_1)+\l X{\tilde \P}_1\P_1, \label{eq:ftermsuper}
\eeq
where $\P_i,\tilde{\P}_i$ ($i=1,2$) are the messenger fields, and $X$ the SUSY breaking singlet field.
${\tilde \P}_i$ and $\P_i$ transform in the $n_5 \times \bf 5$ and $n_5 \times \bf \bar{5}$ representation of 
$SU(5)_{\rm GUT}$ respectively, where $2n_5$ is the number of the messengers.
This model was studied in some of early works on gauge mediation~\cite{Dine:1981za}, and
also studied in Ref.~\cite{Izawa:1997gs} with an implementation of dynamical SUSY breaking, 
and later found as an effective theory describing direct mediation models~\cite{Kitano:2006xg,Csaki:2006wi}
in the Intriligator-Seiberg-Shih (ISS) metastable vacuum~\cite{Intriligator:2006dd}.
The SUSY breaking vacua with $\P=\tilde{\P}=0$ are stable, as long as $|\l F|<|M|^2$. The field $X$ spans a moduli space at tree level, and 
we assume that the VEV of $X$ is stabilized at a nonzero value $\vev{X} \neq 0$ to break the  R-symmetry of the theory.
We can take all parameters to be real without loss of generality.

According to Ref.~\cite{Izawa:1997gs}, by appropriately choosing $\vev{X}$ to maximize the gaugino masses (the maximization occurs at $\l\vev{X}\simeq M$),
 we obtain
\beq
M_a\simeq n_5\frac{g_a^2}{16\pi^2} \left(0.1\frac{(\l F)^3}{M^5}+{\cal O}((\l F)^5/M^{9})\right). \label{eq:stableftermgaugino}
\eeq
Note that there is no term of order $\l F/M$. This is a consequence of the stability of the SUSY breaking vacuum in the above model~\cite{Komargodski:2009jf},
as discussed in the previous subsection.
Furthermore, there is an additional suppression by the numerical factor $0.1$ in the first term of the above equation.

On the other hand, the sfermion masses are given by
\beq
m_{\f_i}^2 \simeq 2n_5 \sum_{a=1,2,3} \left(\frac{g_a^2}{16\pi^2} \right)^2 C_a(i) \left(\left(\frac{\l F}{M}\right)^2+{\cal O}((\l F)^4/M^6)\right),\label{eq:stableftermsfermion}
\eeq
as long as $\l\vev{X}\lsim M$. There are no suppression factors which are present in the gaugino masses. If $\sqrt{\l F}/M$ is small compared with $1$,
the ratio of e.g. the Bino mass to the right handed selectron mass is given by
\beq
\frac{M_{\tilde B}}{m_{\tilde e_R}} \simeq 0.1\sqrt{n_5}\left(\frac{\sqrt{\l F}}{M}\right)^{4}.
\label{Reg2}
\eeq
For the perturbative GUT unification to be successful, $n_5$ cannot be too large.
Thus, without tuning $\l F \simeq M^2$, the gaugino masses are suppressed compared with the
sfermion masses.

The gravitino mass is given by
\beq
m_{3/2}= \frac{F}{\sqrt{3}M_{Pl}} \simeq 35~\EV \left(\frac{1}{\l}\right)\left(\frac{2}{n_5}\right)^2 \left(\frac{M}{\sqrt{\l F}}\right)^{10}\left(\frac{M_{\tilde B}}{100~{\rm GeV}}\right)^2.
\eeq
More detailed study of Ref.~\cite{Sato:2009dk} shows that it is difficult to achieve 
very light gravitino mass $m_{3/2} \lsim16~\EV$ in the model, even if we tune $\l F \simeq M^2$. 
Thus we have to consider a heavy enough gravitino mass to avoid cosmological gravitino problems (see section~\ref{subsec:4.1}). 
For that, it is necessary for $\l F/M^2$ and/or $\l$ to be small. 
Notice that the smallness of $\l F/M^2$ leads to the splitting between the gaugino and sfermion masses. 
For example, by taking $\l \simeq 1$, $n_5=2$, $M_{\tilde B} \simeq 100~{\rm GeV}$ and $M/\sqrt{\l F}\simeq 3$, we obtain 
$m_{3/2} \simeq 2~{\rm MeV}$ and $m_{\tilde e}/M_{\tilde B} \simeq 600$.

The violation of the GUT relation of the gaugino masses can be calculated as in section~\ref{sec:GUTtukinuke}, if we assume that the effective superpotential
Eq~(\ref{eq:ftermsuper}) is valid up to the GUT scale. 
In the parametrization of that section,
the present model has 
$p=2$ and $N_5=2n_5$. Then it may be difficult to achieve the gluino NLSP, unless the splitting between the gaugino and sfermion masses is so large 
that the contributions (\ref{eq:higgs1}-\ref{eq:higgs3}) become important.

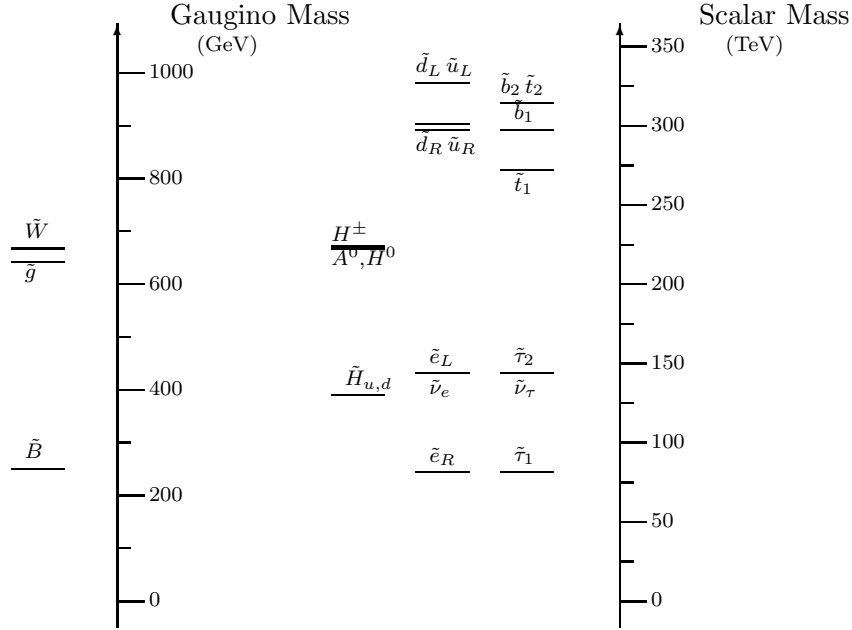
\begin{figure}[htbp]
\begin{center}
\scalebox{1}{
\begin{picture}(280,228.504)(0,0)
\put(260,218.504){\small Scalar Mass }
\put(270,208.504){$\scriptstyle {\rm (TeV)}$ }
\put(230,-10){\vector(0,1){228.504}}
\put(230,0){\line(1,0){10}}
\put(242,-2){$\scriptstyle 0$}
\put(230,30){\line(1,0){10}}
\put(242,28){$\scriptstyle 50$}
\put(230,60){\line(1,0){10}}
\put(242,58){$\scriptstyle 100$}
\put(230,90){\line(1,0){10}}
\put(242,88){$\scriptstyle 150$}
\put(230,120){\line(1,0){10}}
\put(242,118){$\scriptstyle 200$}
\put(230,150){\line(1,0){10}}
\put(242,148){$\scriptstyle 250$}
\put(230,180){\line(1,0){10}}
\put(242,178){$\scriptstyle 300$}
\put(230,210){\line(1,0){10}}
\put(242,208){$\scriptstyle 350$}
\put(230,0){\line(1,0){5}}
\put(230,15){\line(1,0){5}}
\put(230,30){\line(1,0){5}}
\put(230,45){\line(1,0){5}}
\put(230,60){\line(1,0){5}}
\put(230,75){\line(1,0){5}}
\put(230,90){\line(1,0){5}}
\put(230,105){\line(1,0){5}}
\put(230,120){\line(1,0){5}}
\put(230,135){\line(1,0){5}}
\put(230,150){\line(1,0){5}}
\put(230,165){\line(1,0){5}}
\put(230,180){\line(1,0){5}}
\put(230,195){\line(1,0){5}}
\put(121,134.361){\line(1,0){20}}
\put(121,133.483){\line(1,0){20}}
\put(121,133.483){\line(1,0){20}}
 \put(121,136.483){$\scriptstyle H^{\pm}$}
 \put(121,127.361){$\scriptstyle A^0,H^0$}
\put(0,50.1969){\line(1,0){20}}
\put(5.,54.1969){$\scriptstyle {\tilde{B}}$}
\put(0,133.512){\line(1,0){20}}
\put(5.,137.512){$\scriptstyle {\tilde{W}}$}
\put(0,128.305){\line(1,0){20}}
\put(5,122.305){$\scriptstyle {\tilde g}$}
\put(60,218.504){\small Gaugino Mass }
\put(70,208.504){$\scriptstyle {\rm (GeV)}$ }
\put(40,-10){\vector(0,1){228.504}}
\put(40,0){\line(1,0){10}}
\put(52,-2){$\scriptstyle 0$}
\put(40,40){\line(1,0){10}}
\put(52,38){$\scriptstyle 200$}
\put(40,80){\line(1,0){10}}
\put(52,78){$\scriptstyle 400$}
\put(40,120){\line(1,0){10}}
\put(52,118){$\scriptstyle 600$}
\put(40,160){\line(1,0){10}}
\put(52,158){$\scriptstyle 800$}
\put(40,200){\line(1,0){10}}
\put(52,198){$\scriptstyle 1000$}
\put(40,0){\line(1,0){5}}
\put(40,20){\line(1,0){5}}
\put(40,40){\line(1,0){5}}
\put(40,60){\line(1,0){5}}
\put(40,80){\line(1,0){5}}
\put(40,100){\line(1,0){5}}
\put(40,120){\line(1,0){5}}
\put(40,140){\line(1,0){5}}
\put(40,160){\line(1,0){5}}
\put(40,180){\line(1,0){5}}
\put(121,77.9424){\line(1,0){20}}
\put(126.,81.9424){$\scriptstyle {\tilde{H}_{u,d}}$}
\put(153,196.237){\line(1,0){20}}
\put(153,196.237){\line(1,0){20}}
\put(153,200.237){$\scriptstyle \tilde d_L\,\tilde u_L$}
\put(153,178.465){\line(1,0){20}}
\put(153,180.747){\line(1,0){20}}
\put(153,170.465){$\scriptstyle \tilde d_R\,\tilde u_R$}
\put(153,86.3575){\line(1,0){20}}
\put(153,86.3575){\line(1,0){20}}
\put(158,78.3575){$\scriptstyle \tilde \nu_e$}
\put(158,90.3575){$\scriptstyle \tilde e_L$}
\put(153,48.8947){\line(1,0){20}}
\put(158,52.8947){$\scriptstyle \tilde e_R $}
\put(185,163.28){\line(1,0){20}}
\put(190,155.28){$\scriptstyle \tilde t_1$}
\put(185,188.504){\line(1,0){20}}
\put(185,178.447){\line(1,0){20}}
\put(190,182.447){$\scriptstyle \tilde b_1$}
\put(185,188.504){\line(1,0){20}}
\put(185,192.504){$\scriptstyle \tilde b_2\,\tilde t_2$}
\put(185,48.925){\line(1,0){20}}
\put(190,52.925){$\scriptstyle \tilde \tau_1$}
\put(185,86.3786){\line(1,0){20}}
\put(190,90.3786){$\scriptstyle \tilde \tau_2$}
\put(185,86.3786){\line(1,0){20}}
\put(190,78.3786){$\scriptstyle \tilde \nu_{\tau}$}
\end{picture} }
\end{center}
\caption{Low-energy MSSM spectrum in $F$-term GMSB with stable vacuum.
We set $\lambda=1$, $F=8.3\times 10^{15}~\GeV^2$, $M_\ell=2.7\times 10^{8}$ GeV and $\tan\beta=5$.
}
\label{fig:R-spe}
\end{figure}

In Fig.~\ref{fig:R-spe}, we show an example of low-energy MSSM spectrum of this model. In the example, the gluino is lighter that the Wino but is not
the NLSP. If the splitting between the gaugino and sfermion is much larger, the gluino NLSP is also possible.

\bigskip
Finally let us comment on the case where the superpotential Eq.~(\ref{eq:ftermsuper}) is the effective superpotential in the ISS model and the dynamical scale of the ISS model
is much smaller than the GUT scale. In this case, the argument of section~\ref{sec:GUTtukinuke} breaks down and the Wino NLSP may be possible.
In the ISS model, the SUSY is broken in massive supersymmetric QCD (SQCD), and we gauge a subgroup of the flavor symmetry of the model to obtain
the direct gauge mediation model.
At high energies, the electric description of the SQCD is better. The quarks $Q,\tilde{Q}$ of the SQCD are charged under
the SM gauge groups and have superpotential terms
\beq
W_{\rm ele}=m_d Q_d\tilde{Q_d}+m_{\ell}Q_\ell \tilde{Q_\ell}+\cdots,
\eeq
where $Q_d,\tilde{Q_d}$ are charged under $SU(3)_C$, $Q_\ell, \tilde{Q_\ell}$ are charged under $SU(2)_L$, and dots denote mass terms for quarks which 
are not charged under the SM gauge group.
After the confinement of the SQCD, $Q\tilde{Q}$ become mesons.
We denote the mesons by $X$, i.e. 
\beq
Q_d\tilde{Q}_d&=&\sum_{a=SU(3)_C~{\rm index}}Q_d^a\tilde{Q}_d^a \equiv \sqrt{3}\L X_d, \\
Q_\ell \tilde{Q}_\ell&=&\sum_{a=SU(2)_L~{\rm index}}Q_\ell^a\tilde{Q}_\ell^a \equiv \sqrt{2}\L X_{\ell}, 
\eeq
where $\L$ is the dynamical scale of the SQCD. 
The effective superpotential is now given by
\beq
W_{\rm mag} =\sqrt{3}m_d\L X_d+\sqrt{2}m_{\ell}\L X_{\ell}+\cdots \eeq 
where dots denote terms including dual quarks. At low energies, $X_d$ and $X_\ell$ break SUSY, 
and they couple to the $\bf 3$ messengers and $\bf 2$ messengers, respectively. Finally, the part of the effective potential relevant to gauge mediation
is given by,
\beq
W&=&\left[\sqrt{3}m_d \L X_d+\l'_{d} v({\tilde \P}_{d1}\P_{d2}+{\tilde \P}_{d2}\P_{d1})+\frac{\l_d}{\sqrt{3}} X_d{\tilde \P}_{d1}\P_{d1}\right]  \nonumber \\ &&+
\left[\sqrt{2}m_{\ell}\L X_\ell +\l'_{\ell} v({\tilde \P}_{\ell 1}\P_{\ell 2}+{\tilde \P}_{\ell 2}\P_{\ell 1})+\frac{\l_\ell}{\sqrt{2}} X_\ell{\tilde \P}_{\ell 1}\P_{\ell 1}\right],
\eeq
where $v$ is a VEV of some low energy field, 
and $\l_i,\l'_i~~(i=d,\ell)$ are the Yukawa couplings between the dual quarks and mesons which appear in the Seiberg duality.
See e.g. Ref.~\cite{Sato:2009dk} for a detailed discussion on these matters.

The renormalization group flow of the parameters is as follows. At the GUT scale~\footnote{
The model suffers from the Landau pole problem of the SM gauge couplings in some parameter region, especially in low-scale gauge mediation. 
We neglect the problem for simplicity.}, 
we assume that $m_d=m_\ell$.
Then, the RG evolution makes $m_i\L~~(i=d,\ell)$ different at the messenger scale $M_{\rm mess}$,
\beq
\frac{m_d\L}{m_\ell \L} \simeq \exp \left[-\int^{M_{\rm GUT}} _\L \frac{d\mu}{\mu} (\g_{Q_d}+\g_{\tilde{Q}_d}- \g_{Q_\ell}-\g_{\tilde{Q}_\ell})-\int_{M_{\rm mess}} ^\L \frac{d\mu}{\mu} (\g_{X_d}- \g_{X_\ell}) \right],
\eeq
where the anomalous dimension of a chiral field $\F$ is denoted by $\g_\F$.
Since the Yukawa couplings $\l_i,\l'_i~~(i=d,\ell)$ are produced by the confinement dynamics,
they are presumably be the same at the dynamical scale $\L$, $\l_d=\l_\ell=\l'_d=\l'_\ell$.
Then, at the messenger scale, the ratio of these parameters are given by
\beq
\frac{\l_d}{\l_\ell}&\simeq&\exp \left[-\int_{M_{\rm mess}} ^\L \frac{d\mu}{\mu} (\g_{\P_{d1}}+\g_{\tilde{\P}_{d1}}+\g_{X_d}-\g_{\P_{\ell1}}-\g_{\tilde{\P}_{\ell1}}- \g_{X_\ell}) \right], \\
\frac{\l'_dv}{\l'_\ell v}&\simeq&\exp \left[-\int_{M_{\rm mess}} ^\L \frac{d\mu}{\mu} (\g_{\P_{d2}}+\g_{\tilde{\P}_{d1}}-\g_{\P_{\ell2}}-\g_{\tilde{\P}_{\ell1}}) \right].
\eeq
where it should be noted that $\g_{\P_{i2}}+\g_{\tilde{\P}_{i1}}=\g_{\P_{i1}}+\g_{\tilde{\P}_{i2}}~~(i=d,\ell)$.

The gaugino masses are given by (recall the expression in Eq.~\eqref{eq:stableftermgaugino})
\beq
M_1&\propto&\frac{\alpha_1}{4\pi}\left(\frac{2}{5}r' + \frac{3}{5}\right),\\
M_2&\propto&\frac{\alpha_2}{4\pi},\\
M_3&\propto&\frac{\alpha_3}{4\pi}r' ,
\eeq
where $r'$ is given by
\beq
r' \simeq \left(\frac{\l_d}{\l_\ell}\right)^3\left(\frac{m_d\L}{m_\ell \L}\right)^3\left(\frac{\l'_d v}{\l'_\ell v}\right)^{-5}.
\eeq
Here we have chosen the VEVs $\vev{X_i}$ to maximize the gaugino masses and used Eq.~(\ref{eq:stableftermgaugino}).
For example, let us consider the extreme case that $\L \sim M_{\rm mess}$. In this case, $r'$ is given by
\beq
r' &\simeq& \exp \left[-3\int^{M_{\rm GUT}} _{M_{\rm mess}}\frac{d\mu}{\mu} (\g_{Q_d}+\g_{\tilde{Q}_d}- \g_{Q_\ell}-\g_{\tilde{Q}_\ell}) \right] \nonumber \\
&\simeq& \exp \left[ 3\int^{M_{\rm GUT}} _{M_{\rm mess}}\left(\frac{8}{3}\frac{\a_3}{2\pi}+\frac{2}{15}\frac{\a_1}{2\pi}-\frac{3}{2}\frac{\a_2}{2\pi}-\frac{3}{10}\frac{\a_1}{2\pi}\right)\right],
\eeq
where in the second line we have used the 1-loop approximation to the anomalous dimensions.
In this case, the gluino and Bino become heavier than in the case of $r'=1$, and the Wino NLSP may be possible if the Higgs-Higgsino threshold effect is
small.

\subsection{Deformed $F$-term Gauge Mediation}
In direct gauge mediation in the ISS vacuum,
it is also possible to deform the model to generate the gaugino masses at the leading order of 
the SUSY breaking scale~\cite{Kitano:2006xg} (see also Ref.~\cite{Zur:2008zg}). 
Let us consider the following deformation of the superpotential Eq.~(\ref{eq:ftermsuper}),
\beq
W &=& -F^\dagger X+M({\tilde \P}_1\P_2+{\tilde \P}_2\P_1)+\l X{\tilde \P}_1\P_1+m'{\tilde \P}_2\P_2 \nonumber \\
&=&-F^\dagger X+
\left(\begin{array}{cc}
{\tilde \P}_1&{\tilde \P}_2
\end{array}\right)
\left(\begin{array}{cc}
\l X &M \\
M&m'
\end{array}\right)
\left(\begin{array}{c}
\P_1 \\ \P_2
\end{array}\right).
\eeq
We assume that the deformation mass $m'$ and the VEV $\vev{X}$ is smaller than $M$ as in Ref.~\cite{Kitano:2006xg}.
Then, the gaugino masses are given by
\beq
M_a\simeq n_5\frac{g_a^2}{16\pi^2}\left( \e \frac{\l F}{M} + \cdots \right),
\eeq
where we have defined the deformation parameter $\e \equiv m'/M$. The dots denote terms suppressed by $\vev{X}/M$ and $\l F/M^2$. The scalar masses are the same as
in Eq.~(\ref{eq:stableftermsfermion}), aside from small corrections suppressed by $\e$.
The ratio of the Bino mass to the right handed selectron mass is given by
\beq
\frac{M_{\tilde B}}{m_{\tilde e_R}} \simeq \e\sqrt{n_5}.
\eeq
This ratio is not suppressed by $\l F/M^2$ or numerical factors, since the gaugino masses are now generated at the leading order of the SUSY breaking scale.
However, the ratio depends on the deformation parameter $\e$ which is smaller than 1.

There is also another type of direct mediation in the ISS model, by using ``uplifted metastable vacua''~\cite{Giveon:2009yu}. 
In that case, there are minimal-gauge-mediation-like contribution to the MSSM spectrum, as in the model of Ref.~\cite{Cheung:2007es}. We do not discuss these models in this paper.

\subsection{$D$-term Gauge Mediation}\label{subsec:D-term}
In Ref.~\cite{Nakayama:2007cf}, a model of gauge mediation is proposed where the dominant source of the SUSY breaking comes from the Fayet-Iliopoulos (FI)
$D$-term. The messenger fields of the model, $\P$ and $\tilde{\P}$, transform under the representation $({\bf N},+1,\bf{5})$ and $(\bar{\bf N},-1,\bar{\bf{5}})$
of $SU(N)\times U(1)_D \times SU(5)_{\rm GUT}$ respectively, where $SU(N)$ and $U(1)_D$ are the hidden sector gauge groups. 
SUSY is broken by the FI term~\footnote{
The model with the constant FI term cannot be consistently coupled to supergravity~\cite{Komargodski:2009pc},
so one has to consider a mechanism to generate the $D$-term dynamically. See Ref.~\cite{Nakayama:2007cf} for a string theory realization. 
Ref.~\cite{Seiberg:2008qj,Elvang:2009gk} can be taken as a field theory realization. Also, the model does not have a messenger parity~\cite{Dimopoulos:1996ig}
and a dangerous $U(1)_Y$ $D$-term is generated. We can avoid this problem by e.g. introducing another pair of messengers with opposite $U(1)_D$ charge.}
of $U(1)_D$, and R symmetry is broken by the gaugino condensation of $SU(N)$. 
The superpotential of the messenger fields is given by a simple mass term,
\beq
W=M\P\tilde{\P}.
\eeq

The order of the sfermion and gaugino masses is estimated by effective operator analysis. The sfermion masses are generated by 2-loop diagrams
and is given by the operator 
\beq
\int d^4\h \left(\frac{g^2_{\rm SM}}{16\pi^2}\right)^2 \frac{|W_{U(1)_D}^2|^2}{M^6} \f^\dagger \f,
\eeq
where $W_{U(1)_D}=D\h+\cdots$ is the field strength of the $U(1)_D$ gauge field, 
$\f$ is an MSSM field, and $g_{\rm SM}$ collectively denotes the SM gauge couplings. 
This operator generates sfermion masses of order
\beq
m^2_{\f} \sim \left(\frac{g^2_{\rm SM}}{16\pi^2}\right)^2 \frac{D^4}{M^6}. \label{eq:Dtermscalar}
\eeq
By explicit loop computation it was shown in Ref.~\cite{Nakayama:2007cf} that the sfermion masses squared are negative if perturbative calculation is reliable, Thus, in Ref.~\cite{Nakayama:2007cf} it is assumed that
the coupling of $SU(N)$ becomes very strong near the messenger mass scale and the strong interaction makes the sfermion mass squared positive.

The SSM gaugino masses are generated at 1-loop level by the following operator:
\beq
\int d^4\h \left(\frac{1}{16\pi^2}\right)\frac{|W_{U(1)_D}^2|^2}{M^{10}} \left(W^2_{SU(N)}\right)^\dagger W^2_{\rm SM}, 
\eeq
where $W_{SU(N)}$ is the field strength of the $SU(N)$ gauge field, and $W_{\rm SM}$ collectively denotes the SM gauge field strength.
At low energies, gaugino condensation occurs and the VEV of $W^2_{SU(N)}$ is given by the dynamical scale of $SU(N)$.
Then, this operator generates gaugino masses of order
\beq
M_a \sim \frac{g^2_a}{16\pi^2} \frac{\Lambda^3 D^4}{M^{10}}, \label{eq:Dtermgaugino}
\eeq
where $\Lambda^3 \equiv \vev{W_{SU(N)}^2}$. The condition that the $SU(N)$ gauge interaction becomes strong near the messenger mass scale implies that 
\beq
\L \sim M.
\eeq
The ratio of $M_{\tilde B}$ to $m_{\tilde e_R}$ is given by
\beq
\frac{M_{\tilde B}}{m_{\tilde e_R}} \sim \frac{\L^3 D^2}{M^7} \sim \left(\frac{\L}{M}\right)^3\left(\frac{\sqrt{D}}{M}\right)^4.
\eeq

The vacuum energy in global SUSY limit is estimated to be $V \sim D^2$, where we have assumed that the $U(1)_D$ gauge coupling is 
${\cal O}(1)$ and that the $D$-term dominates the vacuum energy.
Then, the gravitino mass is estimated to be
\beq
m_{3/2} \sim \frac{D}{M_{Pl}} \sim 1~\EV \times \left(\frac{M}{\L}\right)^6 \left(\frac{M}{\sqrt{D}}\right)^{14} \left(\frac{M_{\tilde B}}{100~{\rm GeV}}\right)^2.
\eeq
It may be possible in this model to achieve a gravitino mass $m_{3/2} < 16~\EV$, in which case the model suffers from no gravitino problems.
But for a slightly small ratio of $\sqrt{D}/M$, the gravitino becomes heavy and the mass splitting $M_{\tilde B}/m_{\tilde e}$ becomes large.

The GUT relation of the gaugino masses is strongly violated in the present model.
Eq.~(\ref{eq:Dtermgaugino}) shows that $p=10$ in the parametrization 
of section~\ref{sec:GUTtukinuke}, so it is probably easy to obtain the gluino NLSP.

In Fig.~\ref{fig:D-spe}, we show an example of the MSSM mass spectrum in $D$-term GMSB.
Here we assume that the masses of the sparticles are exactly given by Eqs.~(\ref{eq:Dtermscalar}), (\ref{eq:Dtermgaugino})
and that $\Lambda=M$.
The mass splitting is not so large and the Higgs-Higgsino contribution is negligible, but we obtain the gluino NLSP due to the large value of $p$.

\begin{figure}[htbp]
\begin{center}
\scalebox{1}{
\begin{picture}(220,286.671)(0,0)
\put(260,268.504){\small Scalar Mass }
\put(270,258.504){$\scriptstyle {\rm (TeV)}$ }
\put(60,268.504){\small Gaugino Mass }
\put(70,258.504){$\scriptstyle {\rm (GeV)}$ }
\put(230,-10){\vector(0,1){276.671}}
\put(230,0){\line(1,0){10}}
\put(242,-2){$\scriptstyle 0$}
\put(230,60){\line(1,0){10}}
\put(242,58){$\scriptstyle 2$}
\put(230,120){\line(1,0){10}}
\put(242,118){$\scriptstyle 4$}
\put(230,180){\line(1,0){10}}
\put(242,178){$\scriptstyle 6$}
\put(230,240){\line(1,0){10}}
\put(242,238){$\scriptstyle 8$}
\put(230,0){\line(1,0){5}}
\put(230,30){\line(1,0){5}}
\put(230,60){\line(1,0){5}}
\put(230,90){\line(1,0){5}}
\put(230,120){\line(1,0){5}}
\put(230,150){\line(1,0){5}}
\put(230,180){\line(1,0){5}}
\put(230,210){\line(1,0){5}}
\put(121,200.593){\line(1,0){20}}
\put(121,199.267){\line(1,0){20}}
\put(121,199.267){\line(1,0){20}}
 \put(121,202.267){$\scriptstyle H^{\pm}$}
 \put(121,193.593){$\scriptstyle A^0,H^0$}
\put(0,102.263){\line(1,0){20}}
\put(5.,106.263){$\scriptstyle {\tilde{B}}$}
\put(0,208.697){\line(1,0){20}}
\put(5.,212.697){$\scriptstyle {\tilde{W}}$}
\put(0,89.4715){\line(1,0){20}}
\put(5,93.4715){$\scriptstyle {\tilde g}$}
\put(40,-10){\vector(0,1){276.671}}
\put(40,0){\line(1,0){10}}
\put(52,-2){$\scriptstyle 0$}
\put(40,40){\line(1,0){10}}
\put(52,38){$\scriptstyle 200$}
\put(40,80){\line(1,0){10}}
\put(52,78){$\scriptstyle 400$}
\put(40,120){\line(1,0){10}}
\put(52,118){$\scriptstyle 600$}
\put(40,160){\line(1,0){10}}
\put(52,158){$\scriptstyle 800$}
\put(40,200){\line(1,0){10}}
\put(52,198){$\scriptstyle 1000$}
\put(40,240){\line(1,0){10}}
\put(52,238){$\scriptstyle 1200$}
\put(40,0){\line(1,0){5}}
\put(40,20){\line(1,0){5}}
\put(40,40){\line(1,0){5}}
\put(40,60){\line(1,0){5}}
\put(40,80){\line(1,0){5}}
\put(40,100){\line(1,0){5}}
\put(40,120){\line(1,0){5}}
\put(40,140){\line(1,0){5}}
\put(40,160){\line(1,0){5}}
\put(40,180){\line(1,0){5}}
\put(40,200){\line(1,0){5}}
\put(40,220){\line(1,0){5}}
\put(121,46.8016){\line(1,0){20}}
\put(126.,50.8016){$\scriptstyle {\tilde{H}_{u,d}}$}
\put(153,250.651){\line(1,0){20}}
\put(153,250.644){\line(1,0){20}}
\put(153,254.651){$\scriptstyle \tilde d_L\,\tilde u_L$}
\put(153,188.3){\line(1,0){20}}
\put(153,193.441){\line(1,0){20}}
\put(153,200.3){$\scriptstyle \tilde d_R\,\tilde u_R$}
\put(153,171.149){\line(1,0){20}}
\put(153,171.139){\line(1,0){20}}
\put(158,163.139){$\scriptstyle \tilde \nu_e$}
\put(158,175.149){$\scriptstyle \tilde e_L$}
\put(153,75.9812){\line(1,0){20}}
\put(158,79.9812){$\scriptstyle \tilde e_R $}
\put(185,153.727){\line(1,0){20}}
\put(190,145.727){$\scriptstyle \tilde t_1$}
\put(185,236.671){\line(1,0){20}}
\put(185,188.266){\line(1,0){20}}
\put(190,192.266){$\scriptstyle \tilde b_1$}
\put(185,236.644){\line(1,0){20}}
\put(185,240.644){$\scriptstyle \tilde b_2\,\tilde t_2$}
\put(185,75.967){\line(1,0){20}}
\put(190,79.967){$\scriptstyle \tilde \tau_1$}
\put(185,171.136){\line(1,0){20}}
\put(190,175.136){$\scriptstyle \tilde \tau_2$}
\put(185,171.126){\line(1,0){20}}
\put(190,163.126){$\scriptstyle \tilde \nu_{\tau}$}
\end{picture} }
\end{center}
\caption{Low-energy MSSM spectrum in $D$-term GMSB.
We set $\sqrt{D}/M_{\ell}=0.7$, $M_{\ell}=10^{7}~\GeV$, $M_{d}/M_{\ell}=1.3$ and $\tan\beta=2$.}
\label{fig:D-spe}
\end{figure}

\subsection{Summary of Results}

\begin{table}[htbp]
\caption{The mass splitting parameter $R$, the gravitino mass $m_{3/2}$, and the parameter characterizing the GUT relation breaking $p$ for each model
discussed in this section. $R$ and $m_{3/2}$ are evaluated only at the leading order, and these values may become quite different when the MSSM quantum
corrections are taken into account. The meaning of the parameters appearing in the table is as follows. 
In minimal GMSB, $\l$ is the messenger Yukawa coupling and $N_5$ is the messenger number.
In semi-direct GMSB, $h$ is the hidden sector Yukawa coupling,
$h_g=g^2_{SU(2)_{\rm hid}}/8\pi^2$ is the hidden sector $SU(2)_{\rm hid}$ gauge coupling, and $L_V=\log(M_V/M)$ is the log of the ratio of $SU(2)_{\rm hid}$
gauge boson mass to the messenger mass. We have neglected the negative term in Eq.~(\ref{eq:sfermionm}).
In stable $F$-term GMSB, $\l$ is the messenger Yukawa coupling, and $n_5$ is the messenger number divided by two. 
We have assumed that the low energy effective action is valid up to the GUT scale, otherwise the description by $p$ is not valid.
In deformed $F$-term GMSB, the meaning of $\l$ and $n_5$ are the same as in the stable $F$-term GMSB, and $\e$ is a deformation parameter which is
smaller than unity. In $D$-term GMSB, the model is strongly coupled, and we can do only a rough estimation.}

\begin{center}
\begin{tabular}{|c|c|c|c|}
\hline
 & $ \left(\frac{100~{\rm GeV}}{M_{\tilde B}}\right)^2 \cdot m_{3/2}$ &$R$  & $p$    \\ \hline
  minimal & $\gsim 1~{\rm eV}\cdot \left(\frac{1}{N_5^2 \l}\right)$ &$ N_5^{-\frac{1}{2}}$&0 \\ \hline
semi-direct &$\gsim 1~{\rm MeV} \cdot \left(\frac{10^{-4}}{hh_g^{10/3}L_V^{4/3}}\right)$
&$10^3\left( \frac{h h_g L_V}{10^{-2}} \right)^{1 \o 2} \left( \frac{(100~{\rm GeV})^2}{M^2_{\tilde B}}\cdot \frac{m_{3/2}}{100~{\rm MeV}}\right)^{1 \o 2}$&2~{\rm or}~4 \\ \hline
stable $F$-term & $ \gsim 10^2~{\rm eV} \cdot \left(\frac{1}{n_5^2 \l}\right)$ &$ 10 \left( n_5^{3/4} \l \right)^{2 \o 5} 
\left( \frac{(100~{\rm GeV})^2}{M^2_{\tilde B}}\cdot \frac{m_{3/2}}{100~{\rm eV}}\right)^{2 \o 5} $&2 \\ \hline
deformed $F$-term &$\gsim 10~{\rm eV}\cdot \left(\frac{10^{-1}}{n_5^2 \l \e^2}\right)$&$ n_5^{-\frac{1}{2}} \e^{-1}$&0  \\ \hline
$D$-term &$\gsim {\cal O}(1)~{\rm eV}$&${\cal O}(1)\left( \frac{(100~{\rm GeV})^2}{M^2_{\tilde B}}\cdot \frac{m_{3/2}}{1~{\rm eV}}\right)^{2 \o 7}  $&10  \\ \hline
\end{tabular}
\label{table:summ}
\end{center}
\end{table}

In Table~\ref{table:summ}, we give a brief summary of the results discussed in this section \footnote{Let us comment on the difference between our models
and that of Ref.~\cite{Cheung:2007es}. In that paper mass splittings are parametrized by ``effective number of messengers'' in the leading expression 
(with respect to $F/M$) for the gaugino masses, and the splittings do not strongly depend on parameters of the models. 
On the other hand, we are mainly concerned with the case that the leading contribution vanishes and discussing the sub-leading contribution, 
although we have included an example in which the leading contribution does not vanish.
The vanishing of the leading contribution is the source of the violation of the GUT relation of the gaugino masses, even if we impose that at the GUT scale.
However, the cosmological constraint discussed in the next section apply also to the models of Ref.~\cite{Cheung:2007es}.}. 
Here, we have defined the parameter $R$ given by
\beq
R \equiv \frac{m_{\tilde e_R}}{M_{\tilde B}},
\eeq
where $m_{\tilde e_R}$ is the right-handed selectron mass, and $M_{\tilde B}$ is the Bino mass.
$R$ characterizes the splitting between the sfermion and gaugino masses.
$m_{3/2}$ is the gravitino mass, and $p$ is the parameter characterizing the breaking of GUT relation discussed in sec.~\ref{sec:GUTtukinuke}.
The above results are not at all complete and should be taken only as examples, because there are corners of the parameter space where things change
(e.g. tuning the first and second terms in Eq.~(\ref{eq:sfermionm}) to achieve a light sfermion mass)
and also there may be various deformations of the models.
However, the results may be regarded as ``bulk regions of minimal models'' in the whole parameter space.

As we will discuss in the next section, the parameters $R,~m_{3/2}$ and $p$ are quite important for cosmological considerations.
Typically, the Bino may be the NLSP, and the annihilation cross section of the Bino in the early universe sharply depends on $R$. 
If $R$ is large, the annihilation cross section is quite small and the Bino is overproduced in the early universe. 
(Note also that if $R$ is large, the $\mu$-term is typically also large, of the order of the sfermion mass as seen in Eq.~(\ref{eq:muorder}). 
This is an important difference from the split  SUSY~\cite{splitSUSY}, which has a small $\mu$-term.)
The Bino eventually decays to the gravitino, and the decay rate and the gravitino energy density produced by the Bino decay depend on $m_{3/2}$.
The particles produced by the Bino decay (i.e. the gravitino and other SM particles) cause cosmological problems, depending on the value of $m_{3/2}$.
However, if $p$ and/or $R$ is large, the gluino can become the NLSP, and then the situation becomes different from the Bino NLSP case.

\section{Constraints from Cosmology} \label{sec:4}

As we discussed in the previous sections, the low-energy physics of split GMSB at the EW scale will be described by pure Bino, Wino and gluino. 
We assume that the gravitino is the LSP (which is the candidate for the dark matter), and the NLSP is either of these three. We discuss three cases in turn.
Before going to that, let us review the gravitino problem.

\subsection{Gravitino Problem} \label{subsec:4.1}
In the early universe, gravitinos are produced from many kinds of sources.
For example, 
the gravitino is produced from sparticle thermal scattering such as $XY\to\tilde{X}\tilde{G}_{3/2}$
and decay of sparticles $\tilde{X}\to X\tilde{G}_{3/2}$. 
If the gravitino is not thermalized and much lighter than the gluino, its abundance from the thermal scattering
is given~\cite{Bolz:2000fu} as
\beq
\Omega_{3/2}^{\rm sc} h^2 \simeq 0.27 \left(\frac{T_R}{10^{4}~\GeV}\right)
\left(\frac{100~{\rm keV}}{m_{3/2}}\right)
\left(\frac{M_{\tilde{g}}}{1~\TeV}\right)^2, \label{eq:fromscatteing}
\eeq
where $T_R$ is the reheating temperature.
From the decay ~\cite{ArkaniHamed:2004yi},
\beq
\Omega_{3/2}^{\rm dec} h^2 \simeq \sum_{i:{\rm thermalized}} 10^{-2} \times
\left(\frac{M_i}{1~\TeV}\right)^3
\left(\frac{100~{\rm keV}}{m_{3/2}}\right)\times d_i, \label{eq:fromdecay}
\eeq
where $d_i$ is the degree of freedom of sparticle $i$.
If interactions between the gravitino and MSSM matter are strong enough and/or
the reheating temperature $T_R$ is high,
the gravitino is thermalized, and then the abundance is given as
\beq
\Omega_{3/2}^{\rm th} h^2 \simeq 50 \
\left(\frac{m_{3/2}}{100~{\rm keV}}\right).
\eeq
As a result, the gravitino abundance originated from thermal plasma in the early universe is given as
\beq
\Omega_{3/2} h^2 \simeq \min(\Omega_{3/2}^{\rm sc} h^2+\Omega_{3/2}^{\rm dec} h^2,\Omega_{3/2}^{\rm th} h^2).
\eeq
Hereafter, we assume that the reheating temperature $T_R$ is high enough that
at least an SSM low-mass gaugino NLSP is thermalized.
One can see that if the gravitino mass is in the range ${\cal O}(100)~{\rm eV} <m_{3/2}<{\cal O}(10)$ keV, 
the estimated gravitino abundance easily exceeds the current constraint $\Omega h^2 < 0.1$.
It is also known that the bound for a light gravitino mass is given by $m_{3/2}<16~\EV$ from the warm dark matter constraint~\cite{Viel:2005qj}.
It is not so easy to achieve the gravitino mass $m_{3/2}<16~\EV$ (see Table \ref{table:summ}),
so we only consider the case that $m_{3/2}\gsim 100$ keV.

Besides the gravitino production from the thermal plasma,
late-time decay of the NLSP is another source of the gravitino.
In addition, if the lifetime of the NLSP is longer than about $1$ sec, 
its decay has strong impact on cosmology.
In the following subsection, we discuss the effects of the late-time NLSP decay
on cosmology.

\subsection{Bino NLSP}

If the Bino is the NLSP and the gravitino is the LSP,
the Bino is unstable and the gravitino is the candidate for the dark matter.
However, if the Bino is overproduced, the gravitino produced by the Bino decay 
can over-close the Universe.
In addition, late-time decay of the Bino can spoil the success of the 
Big-Bang Nucleosynthesis (BBN) and distort the Cosmic Microwave Background (CMB).
In the Bino NLSP case, it is known that these constrains are very severe \cite{Feng:2004mt}.
In this section, we discuss that these constrains become much severer if $R\gg1$.

First, let us discuss the decay of the Bino.
In the gravitino LSP and the Bino NLSP scenario,
the lifetime of the Bino is written as
\beq
\tau_{\rm NLSP} \simeq 6\times 10^4~ {\rm sec} \left( \frac{m_{3/2}}{1~\GeV} \right)^2 \left(\frac{M_{\rm NLSP}}{100~\GeV} \right)^{-5}. \label{eq:binolifetime}
\eeq
The dominant decay mode of the Bino is $\tilde{B}\to \gamma \tilde{G}_{3/2}$.
If kinematically allowed, the decay modes $\tilde{B}\to Z \tilde{G}_{3/2}$ and 
$\tilde{B}\to h \tilde{G}_{3/2}$ also are open.
In this case, hadronic products from $Z$ and $h$ decay will spoil the BBN.
\footnote{
In this paper, we neglect the mode $\tilde{B}\to h \tilde{G}_{3/2}$, since this process is suppressed in the case of large $\mu$.
}
Even if such modes are closed, there remains hadronic mode such as $\tilde{B}\to \gamma^* \tilde{G}_{3/2}\to q\bar{q} \tilde{G}_{3/2}$.
For example, ${\rm Br}(\tilde{B}\to~{\rm hadrons}) \simeq 0.03,~$ for $m_{\tilde{B}} \lsim 100$ GeV.
\footnote{
Here, we neglect the gravitino mass.
The mode $\tilde{B}\to q\bar{q} \tilde{G}_{3/2}$ has log enhancement term $\log(M_{\tilde{B}}/E_{\rm IR})$, 
where $E_{\rm IR}$ is IR cut-off for the virtual photon propagator.
In this paper we adopt $E_{\rm IR}=m_{q}$.
}
In the case that $m_{\tilde{B}} \gg m_Z$, the branching ratio is about $0.18$.

Next, let us consider the abundance 
\footnote{
Hereafter, we express the relic abundance as the density parameter $\Omega h^2$, as if there is no decay of the NLSP.
The relation between $\Omega_{\rm NLSP} h^2$ and the yield variable $Y_{\rm NLSP}\equiv n_{\rm NLSP}/s$ ($s$ being the entropy density) is given by
$
\Omega_{\rm NLSP} h^2 = 2.82\times 10^8 \times Y_{\rm NLSP}\times (M_{\rm NLSP}/1~\GeV).
$ 
}
of the Bino before its decay.
In general, 
there is strong correlation between
the thermal relic abundance and the annihilation cross section, 
\beq
\Omega h^2 \sim {\cal O}(1) \left(\frac{\left<\sigma v\right>_{\rm Anni}}{10^{-9}~ \GeV ^{-2}} \right)^{-1}.
\eeq

\begin{figure}[htbp]
\begin{center}
\epsfig{file=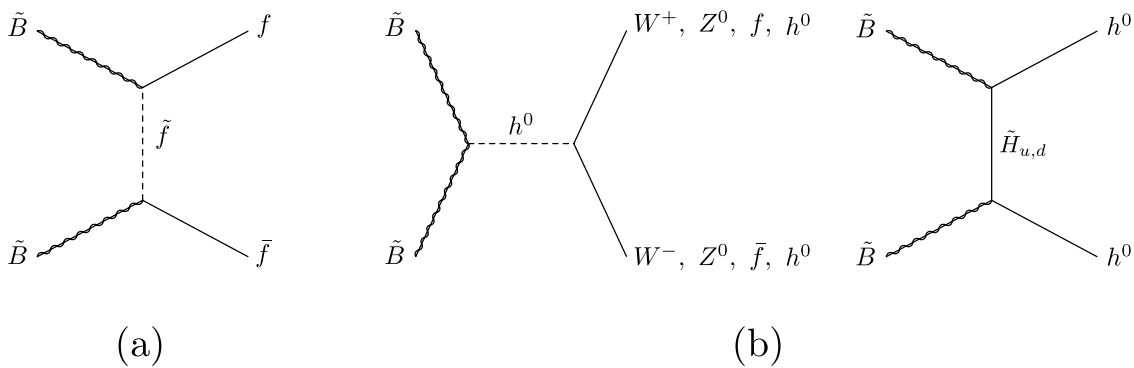 ,scale=.5,clip}\\
\caption{Some of the dominant annihilation processes in split GMSB scenario.
(a): $\left<\sigma v\right> \propto m_s^{-4}$. (b):  $\left<\sigma v\right> \propto \mu^{-2}$.
}
\label{fig:anni}
\end{center}
\end{figure}
The annihilation cross section of the Bino strongly depends on the mass of other sparticles, since the Bino is gauge singlet.
In the case of the Bino NLSP, the annihilation process is dominated by scalar sparticle exchange process (Fig.~\ref{fig:anni})
whose cross section is proportional to $m_{\rm scalar}^{-4}$, or light SM-like Higgs mediated process
whose contribution is approximately proportional to $ \mu^{-2}$.

In the split GMSB scenario, this annihilation process is very suppressed because of very large mass of scalar sparticles
 and Higgsinos. 
Therefore, very large amount of the Bino is produced.
In the case that the sfermion exchange processes are dominant, the Bino abundance is approximately written as
\beq
\Omega^{\rm sfermion}_{\tilde{B}} h^2 \sim {\cal O}(10) \times \left( \frac{100~\GeV}{M_{\tilde{B}}} \right)^2 \left(\frac{M_s}{1~\TeV} \right)^4.
\eeq
In the case that the Higgsino or Higgs scalar exchange processes are dominated,
the Bino abundance is given as
\beq
\Omega^{\rm Higgsino}_{\tilde{B}} h^2 \sim {\cal O}(100) \times \left( \frac{\tan\beta}{10} \right)^2 \left(\frac{\mu}{1~\TeV} \right)^2,
\eeq
for $M_{\tilde{B}}\gsim {\cal O}(100)~\GeV$.
One can see that large mass splitting induces the huge abundance of the Bino.

\begin{figure}[htbp]
\begin{tabular}{cc}
\begin{minipage}{0.5\hsize}
\begin{center}
\epsfig{file=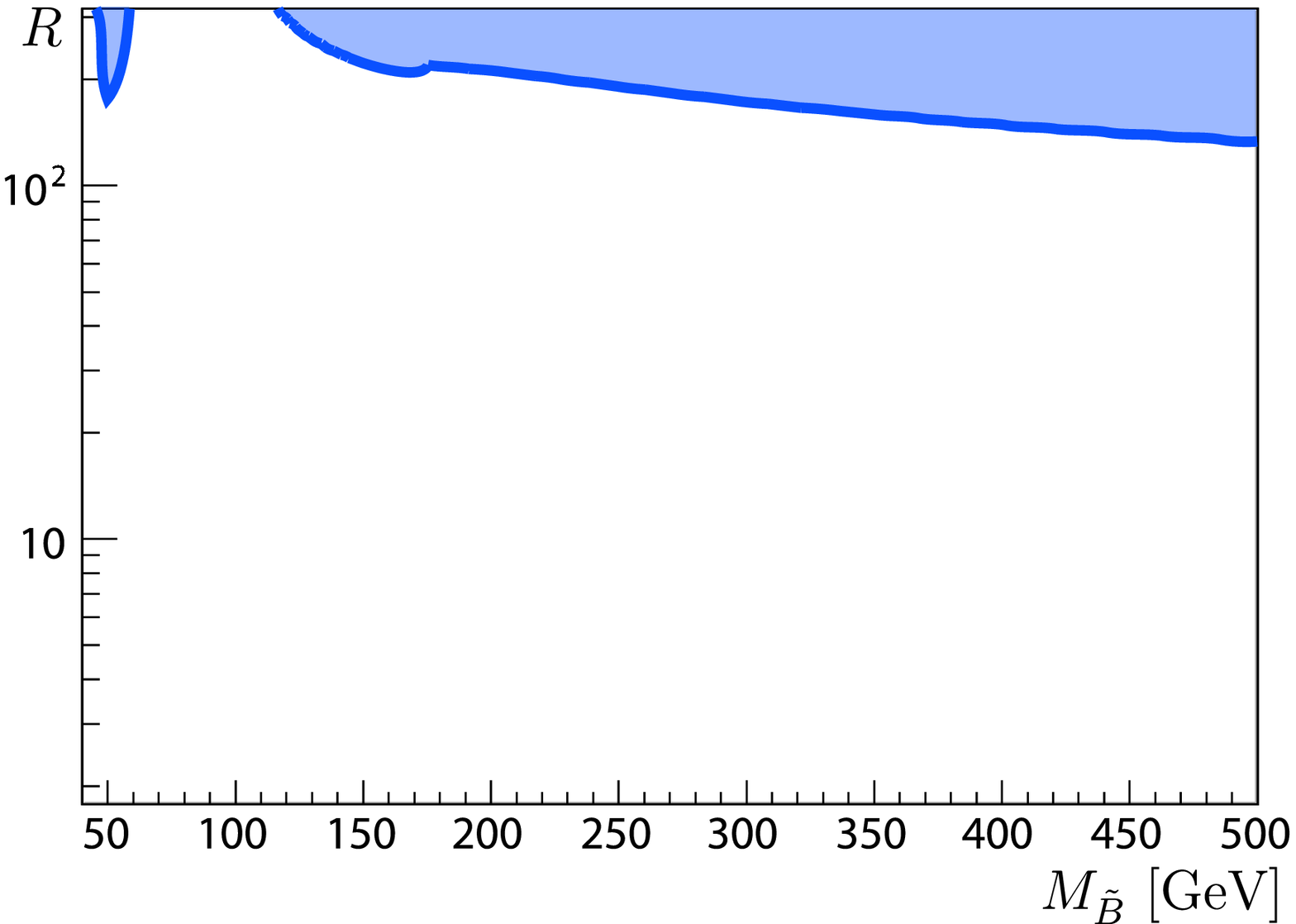 ,scale=.45,clip}
(a) $m_{3/2} = 100$ keV
\end{center}
\end{minipage}
& 
\begin{minipage}{0.5\hsize}
\begin{center}
\epsfig{file=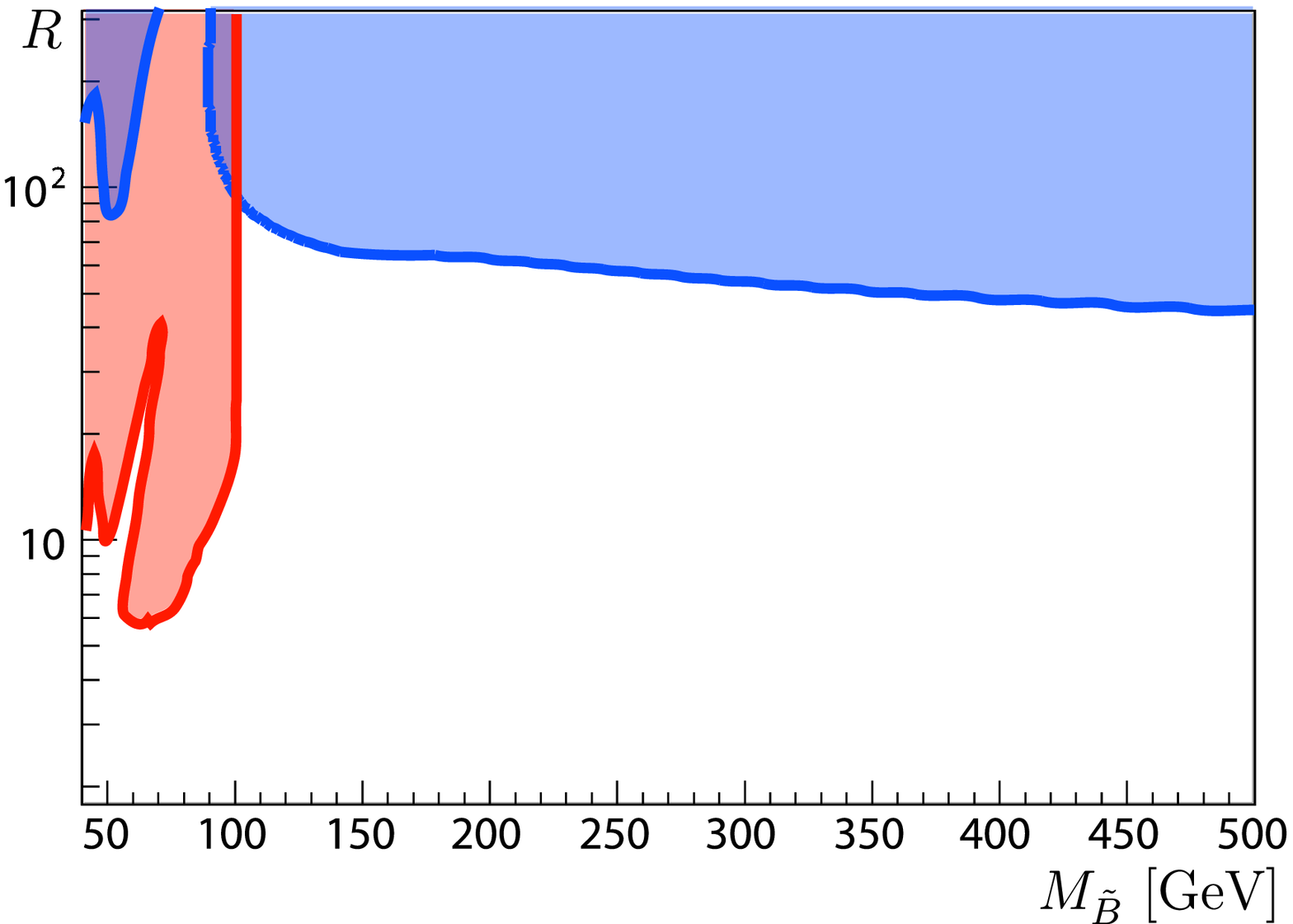 ,scale=.45,clip}
(b) $m_{3/2} = 1$ MeV
\end{center}
\end{minipage}\\
%
 & \\
\begin{minipage}{0.5\hsize}
\begin{center}
\epsfig{file=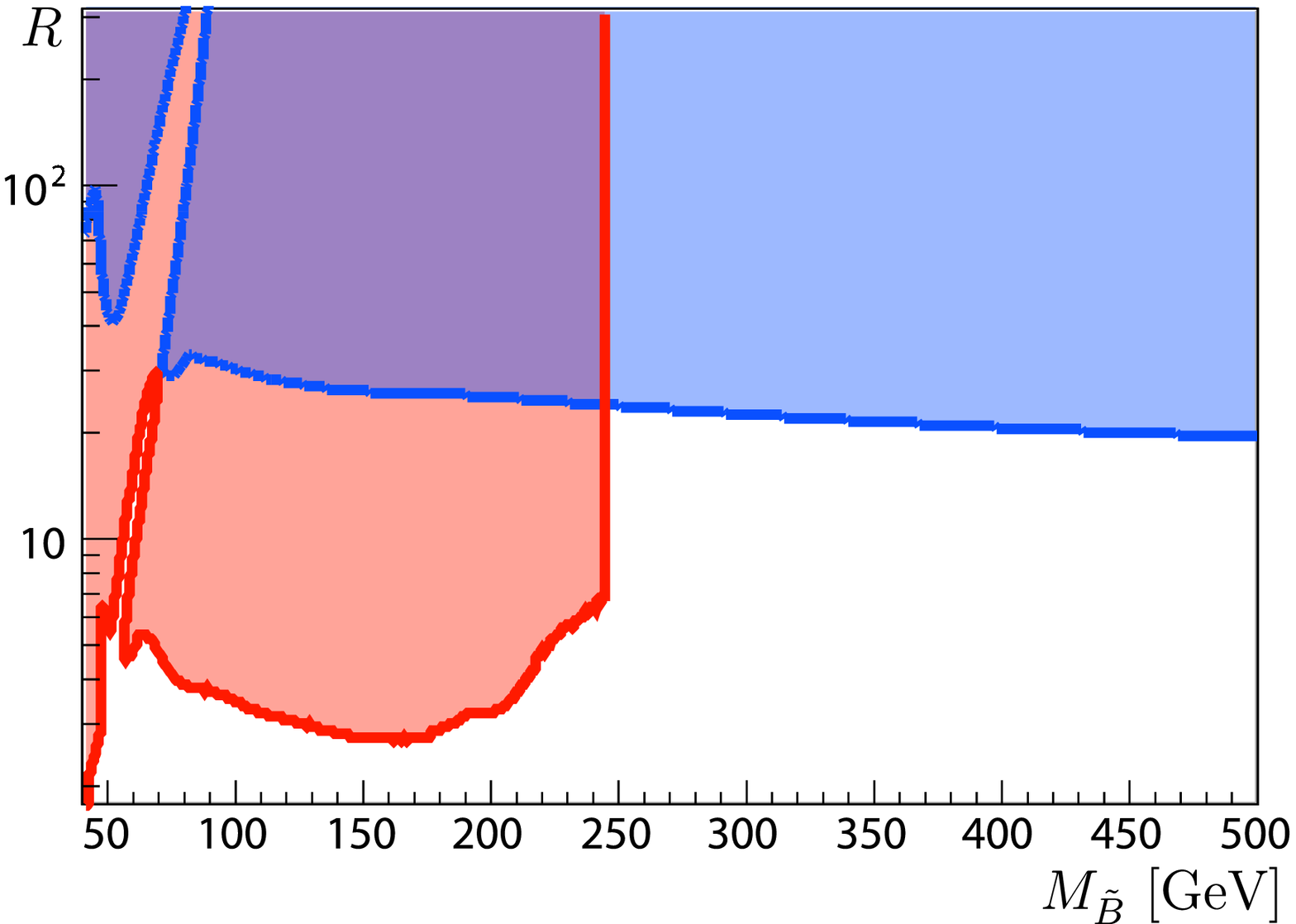 ,scale=.45,clip}
(c) $m_{3/2} = 10$ MeV
\end{center}
\end{minipage}
&
\begin{minipage}{0.5\hsize}
\begin{center}
\epsfig{file=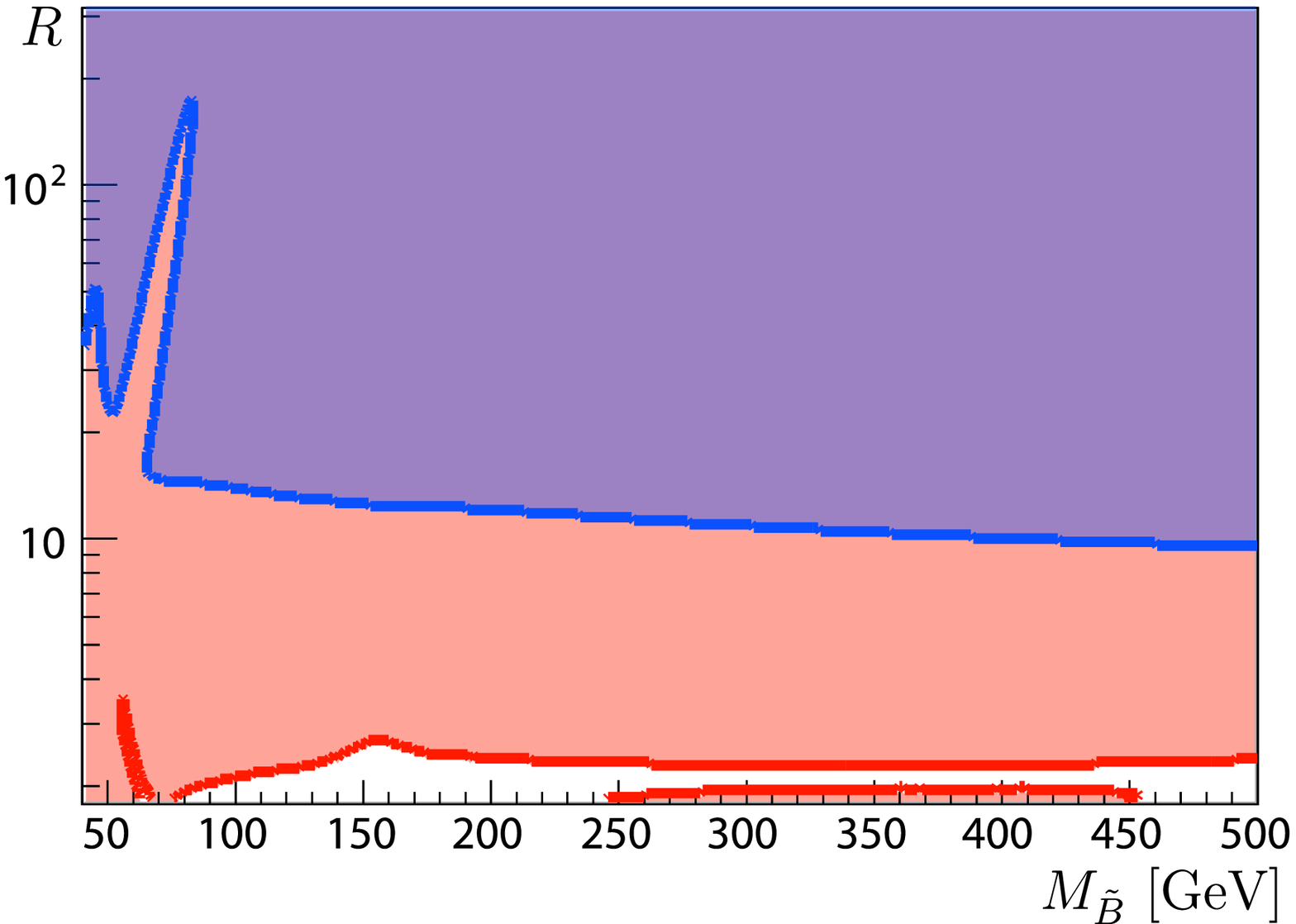 ,scale=.45,clip}
(d) $m_{3/2} = 100$ MeV
\end{center}
\end{minipage}\\
 & \\
\begin{minipage}{0.5\hsize}
\begin{center}
\epsfig{file=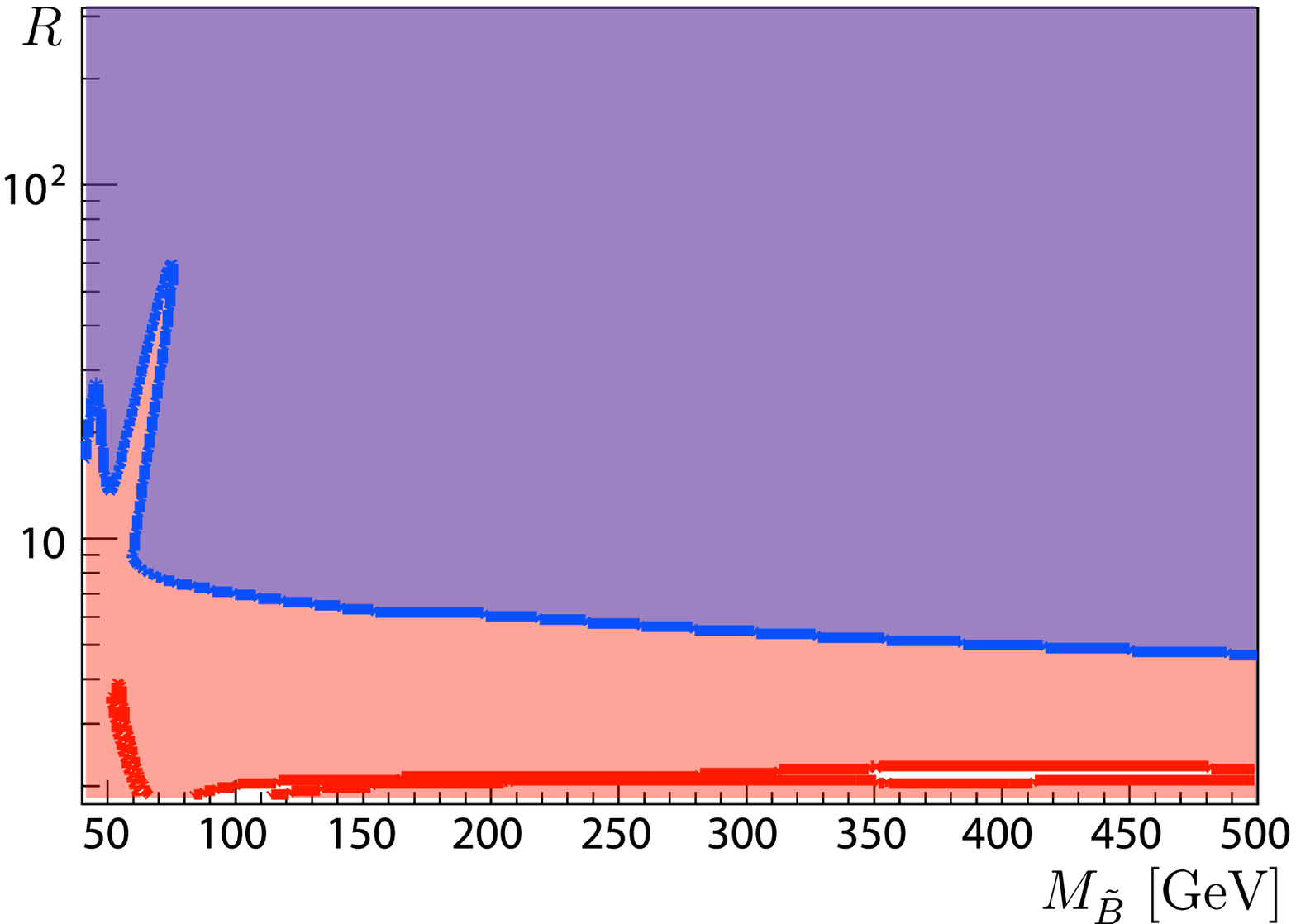 ,scale=.45,clip}
(e) $m_{3/2} = 1$ GeV
\end{center}
\end{minipage}
&
\begin{minipage}{0.5\hsize}
\begin{center}
\epsfig{file=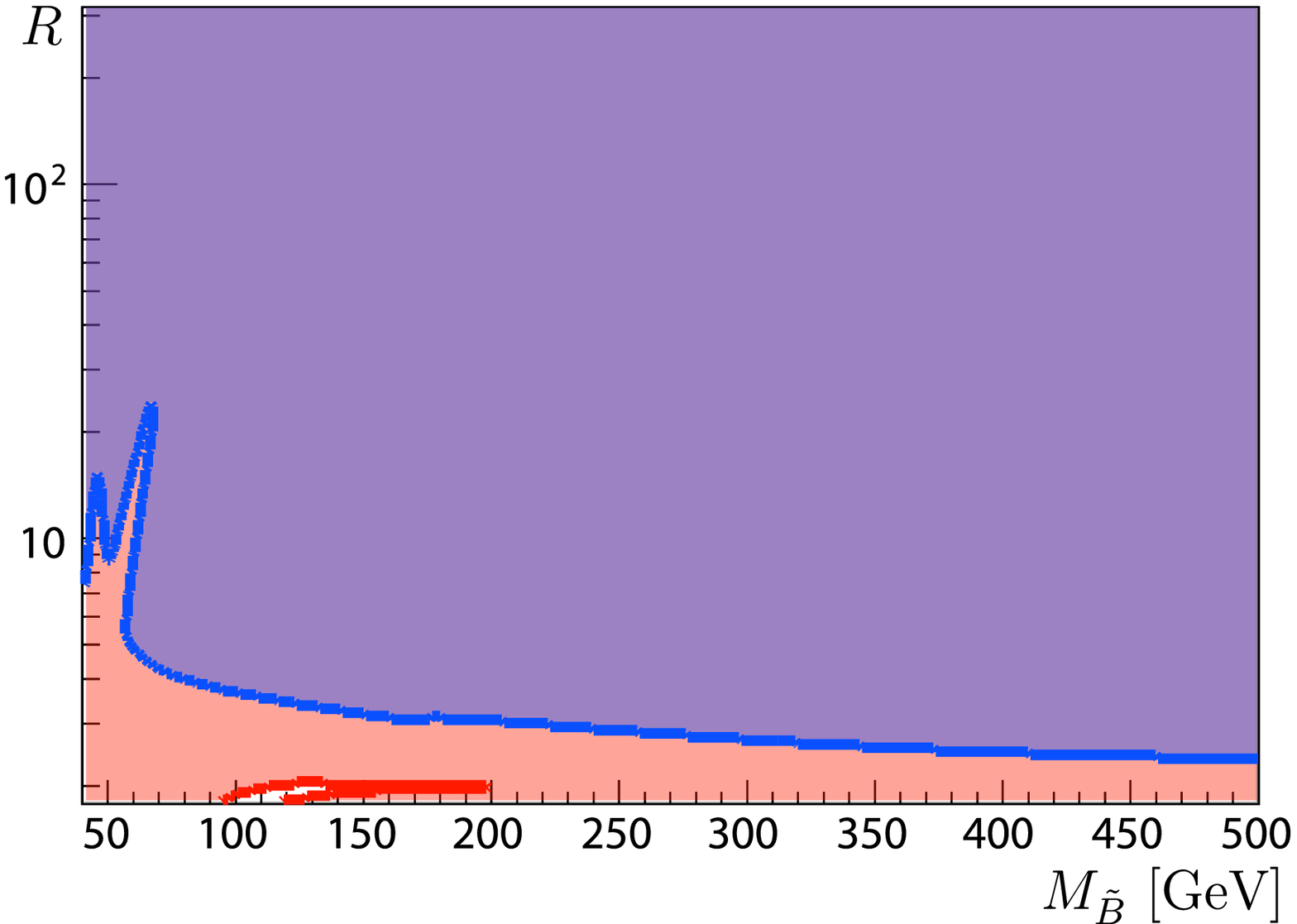 ,scale=.45,clip}
(f) $m_{3/2} = 10$ GeV
\end{center}
\end{minipage}

\end{tabular}
\caption{The constraint on the mass splitting parameter $R$.
The blue region is excluded by the over-close condition and
the red the BBN condition.
}
\label{fig:result}
\end{figure}

Then, let us see the constraint on the mass splitting, following the above discussion.
We have used the program {\verb DarkSUSY }
5.0.2~\cite{Gondolo:2004sc} to estimate the Bino abundance in the BBN period.
We set $M_{\tilde{g}}:M_{\tilde{W}}:M_{\tilde{B}}=6:2:1$, $ \mu = m_{\rm scalar}=R \times M_{\tilde B}$, $A_{\tau,b,t}=0$
and $\tan\beta=10$.
As for the BBN constraint, we follow the result obtained in Ref.~\cite{Jedamzik:2006xz}.
In addition, we impose the following condition,
\beq
\Omega_{3/2} h^2 = \frac{m_{3/2}}{M_{\tilde{B}}}\Omega_{\tilde{B}} h^2<\Omega_{\rm DM} h^2=0.1.
\eeq
In Figs.~\ref{fig:result}, we show the constraint on the mass splitting $R$.
Here we omit the constraints from the CMB distortion, since this constraint is weaker than that of the BBN.
In the region $M_{\tilde{B}} \sim 50$ GeV, the constraint is weak, since  $h^0/Z$ pole effect drastically
reduces the Bino abundance.
As for the gravitino production from the thermal process,
heavy gravitino is favored as seen in Eqs.~(\ref{eq:fromscatteing}) and (\ref{eq:fromdecay}).
On the other hand, as for the Bino decay,
the mass splitting parameter $R$ is strongly constrained in the case of the heavy gravitino.

\subsection{Wino NLSP}
\begin{figure}[htbp]
\begin{center}
\epsfig{file=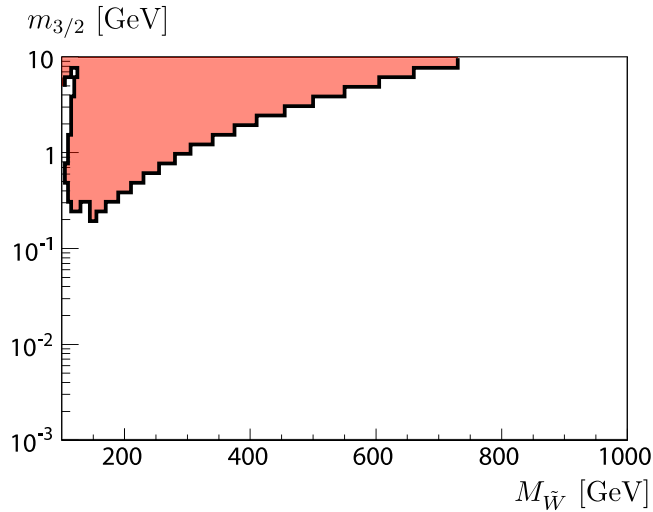 ,scale=.47,clip}
\end{center}
\caption{The constraint for the Wino and gravitino masses from the BBN.
The red region is excluded.
}
\label{fig:winoNLSP}
\end{figure}

In section~\ref{sec:stableftermgmsb}, we discussed that the Wino NLSP may be possible in the split GMSB.
In this case, the story is similar to that of anomaly mediation except that the gravitino mass is much lighter.
Electrically neutral Wino $\tilde{W}^0$ is expected to be lighter than the charged Wino $\tilde{W}^{\pm}$.
Thus only neutral Wino $\tilde{W}^0$ lives in the BBN era.
The lifetime and decay mode of the Wino is similar to the Bino NLSP case.
However, the branching fraction to $Z+\tilde{G}_{3/2}$ is larger than that of the Bino.
In contrast to the Bino NLSP case, the Wino abundance is almost independent of other sparticle masses.
This is because  the Wino is not gauge singlet and the annihilation process $\tilde{W}\tilde{W}\to W^+W^-$ can occur via 
the $SU(2)_L$ interaction, which dominantly determine the Wino abundance. 
Its abundance is approximately given as
\beq
\Omega_{\tilde{W}} h^2 \simeq 2\times 10^{-4} \left( \frac{M_{\tilde{W}}}{100~{\rm GeV}} \right)^2. \label{eq:winoab}
\eeq
The Wino abundance is much smaller than the Bino one.
Thus, the cosmological constraint from the NLSP decay weakens, compared to the Bino NLSP case.
In Fig.~\ref{fig:winoNLSP}, we show the constraint on $M_{\tilde{W}}$ and $m_{3/2}$.
Here, we set sparticle masses other than the Wino to be 10 TeV and $\tan\beta=10$.
As seen in Eq.~(\ref{eq:winoab}), the Wino abundance is so small that
the gravitino abundance from the late-time decay of the Wino cause no problems.
The cosmological constraints come from the BBN ones.
Roughly speaking, if the lifetime of the Wino is smaller than 200 sec,
the BBN constraint can be satisfied.
In other words, if the condition
\beq
m_{3/2}\lsim 6\times10^{-2}~\GeV \left( \frac{M_{\rm NLSP}}{100 ~\GeV} \right)^{5/2}.\label{eq:200sec}
\eeq
is satisfied, there are no cosmological problems.

\subsection{Gluino NLSP}

Let us discuss the gluino NLSP case, which is plausible in the split GMSB scenario as we discussed in the previous sections.
Like the Bino and Wino NLSP, the late time decay of the gluino also has impact on the BBN and CMB.
The lifetime of the gluino is the same as that of the Bino shown in Eq.~(\ref{eq:binolifetime}).
However, estimation of the gluino abundance is complicated because of the strong QCD effects.
There are two stages of the gluino annihilation, which determine the gluino abundance.
One is the age $T\gsim M_{\tilde{g}}/20$.
The gluino abundance during this stage is determined by perturbative annihilation cross section, and the final abundance is given by: \cite{Baer:1998pg}
\beq
\Omega_{\tilde g}^{\rm pert} h^2\sim 10^{-3}\left(\frac{M_{\tilde{g}}}{1~\TeV}\right)^2.
\eeq
Additional annihilation occurs after QCD phase transition.
In this stage, the gluino forms heavy hadronic bound state with quarks or gluons.
The gluino annihilation cross section is enhanced by the strong interaction, and the gluino abundance is reduced.
Although there are several studies for the estimation of its abundance 
\cite{ArkaniHamed:2004fb,Arvanitaki:2005fa,Kang:2006yd,Jacoby:2007nw}, quantitative evaluation of the gluino abundance is difficult because of hadron dynamics.
The papers claim different values for the abundance. 
Among of them, we adopt the result of Ref.~\cite{Kang:2006yd}, whose evaluation is rather small and thus
conservative for the consideration of the cosmological constraint.
It reads
\beq
\Omega_{\tilde{g}}h^2 \sim 10^{-7} \left(\frac{r_{\rm had}}{\GeV^{-1}}\right)^{-2} \left(\frac{T_B}{180~\MeV}\right)^{-3/2}\left(\frac{M_{\tilde{g}}}{1~\TeV}\right)^{3/2}, \label{eq:gab}
\eeq
where $r_{\rm had}$ is effective radius of gluino hadronic bound state and $T_B$ the temperature at which the bound state is formed.
If this is the case, the gluino abundance is low enough that the cosmological constraints discussed previously are
inefficient.
However, authors in Ref.~\cite{Kusakabe:2009jt} discuss that 
long-lived colored particle captured by nuclei have great impacts on the BBN and that the gluino lifetime $\tau_{\tilde g}$ should satisfy
$\tau_{\tilde{g}}\lsim 200$ sec, even if the gluino abundance is
reduced as in Eq.~(\ref{eq:gab}). 
Therefore like the Wino NLSP case, the gravitino mass may satisfy the condition Eq.~(\ref{eq:200sec})
in the gluino NLSP case.

\section{Collider Implication}\label{sec:5}
We have seen that non-minimal GMSB models typically predict a large hierarchy between the gaugino and sfermion masses.
This, of course, is not favorable from the viewpoints of the gauge hierarchy problem and GUT unification. Even if such a SUSY breaking mediation mechanism is realized in nature, there are strong cosmological constraints, especially when the Bino is the NLSP.  We should also keep in mind, however, that there are several possible options for evading the constraints (especially when the Wino or gluino is the NLSP), and it is therefore an interesting problem to study the collider signatures of the split GMSB models, taking the cosmological constraints into account. This is the goal of the present section. 
We will see that the split GMSB models often have rather unconventional signatures from more conventional GMSB scenarios, such as the minimal GMSB.

\begin{figure}[htbp]
\begin{tabular}{cc}
\begin{minipage}{0.5\hsize}
\begin{center}
\epsfig{file=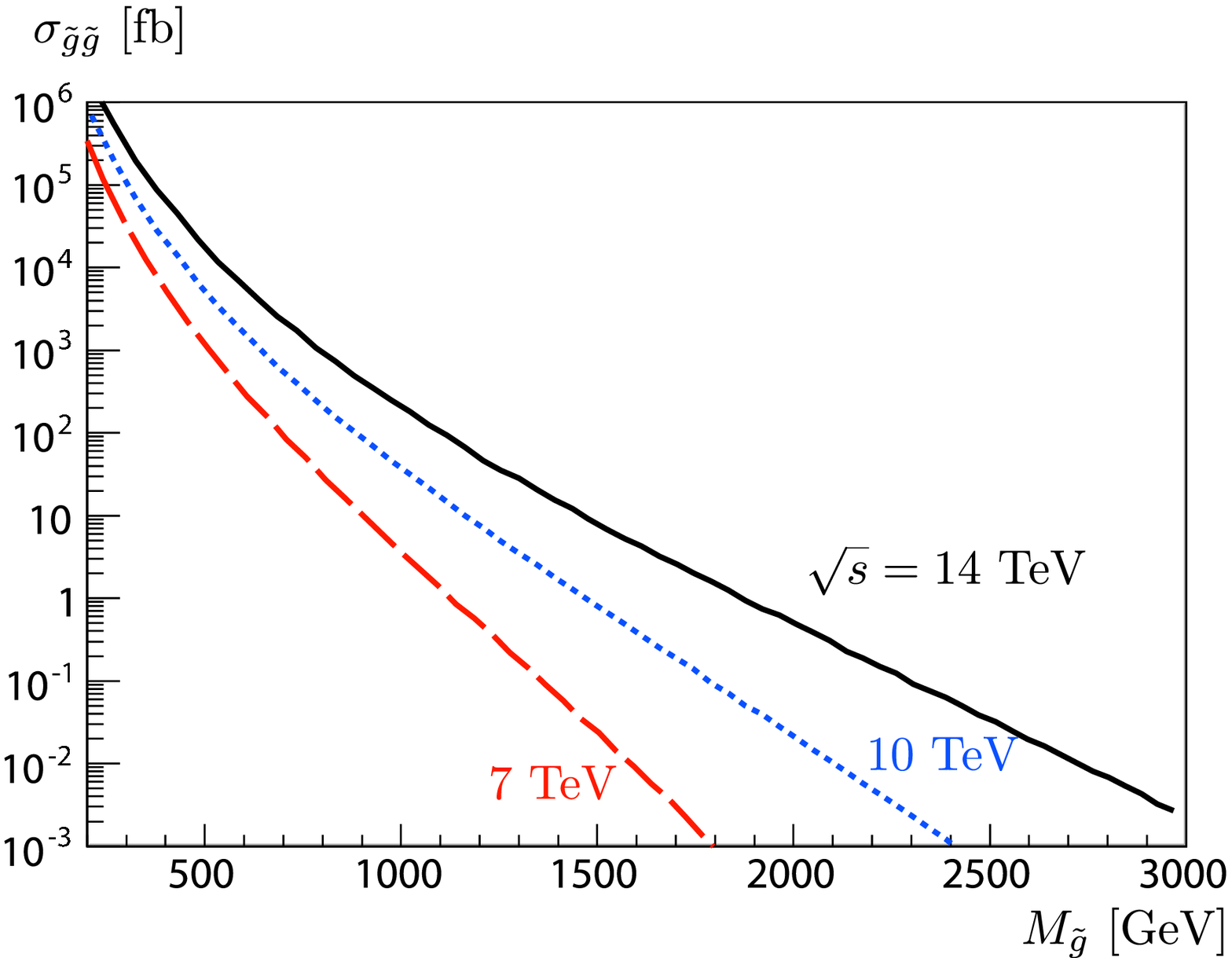 ,scale=.44,clip}
\end{center}
\end{minipage}
\begin{minipage}{0.5\hsize}
\begin{center}
\epsfig{file=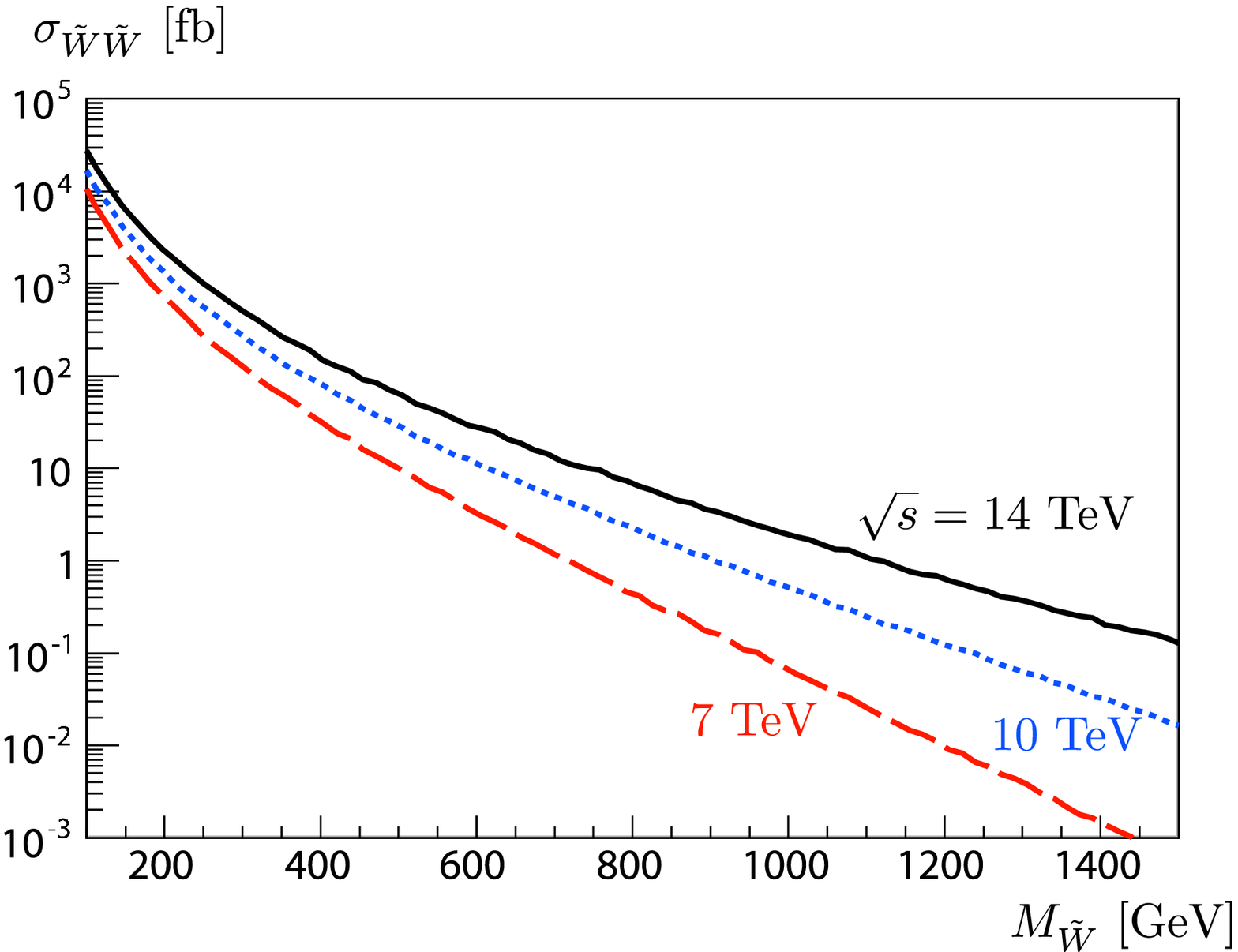 ,scale=.44,clip}
\end{center}
\end{minipage}
\end{tabular}
\caption{Cross section of the gluino (left) and wino (right) pair production at the LHC.
The red and dashed line show the case with $\sqrt{s}=7$ TeV, 
blue and dotted $\sqrt{s}=10$ TeV, and black and solid $\sqrt{s}=14$ TeV.
}
\label{fig:LHCcross}
\end{figure}

Before discussing individual cases, 
let us consider the production of SUSY particles at the LHC.
At the LHC, colored sparticle production is an important channel for SUSY events. In the 
sfermion decoupling limit, the gluino pair production:
\beq
pp\to\tilde{g}\tilde{g}+X
\eeq
is one of the most important production processes, if the gluino is not much heavier than the Wino.
In Figs.~\ref{fig:LHCcross}, we show the leading order cross section for gluino 
and Wino pair production in sfermion decoupling limit.
To estimate the cross section, we have used the program \verb+Pythia 6.4.19+ \cite{Sjostrand:2006za}.
Other sparticles are produced from SUSY cascade decay of the gluino.
For example, if the gaugino masses obey the GUT relation and 
sfermions are much heavier than the gluino, the Bino and Wino are produced through the SUSY cascade decay such as
$\tilde{g}\to \tilde{W}q\bar{q}$ and subsequently $\tilde{W} \to \tilde{B}\ell\bar{\ell}$.
Although the cross section for direct production of the Wino is smaller than that of the gluino, this will be important SUSY production channels if the gluino is heavy.
The Bino production cross section is negligibly small.

\subsubsection*{Bulk Scenario}
First of all, we discuss the LHC signature of GMSB model in which 
MSSM scalar particles and Higgsino are heavy enough to be irrelevant at the LHC without virtual effects,
the gaugino masses obey the GUT relation, and
$m_{3/2}>{\cal O}(10)~\MeV$.
Although this pattern suffers from cosmological problems as seen before,
it is impossible to claim that this mass pattern cannot be realized in nature.
This is because some mechanism such as low-reheating temperature or entropy production,
may solve the cosmological problems caused by the Bino NLSP.

Assuming the gravitino LSP,
the Bino NLSP is not stable and finally decays into a gravitino.
However, its decay length is given as
\beq
c\tau_{\rm NLSP} = 2\times 10^{7}{\rm m} \left(\frac{m_{3/2}}{1~\MeV}\right)^2\left(\frac{m_{\rm NLSP}}{100~\GeV}\right)^{-5}.
\eeq
If the gravitino is much heavier than a few MeV,
then almost all of Bino NLSP decays outside of the detector.
Therefore, in this case, the behavior of the Bino is the same as 
one of traditional neutralino dark matter scenario.
If the gluino is lighter than a few TeV,
multi-jet + missing energy signatures are promising SUSY events.
If the gluino is too heavy to be created at the LHC,  
the production of the Wino is important.
In this case, multi-lepton + missing energy signatures may be 
a precious channel for the SUSY discovery.

\subsubsection*{Light Gravitino and Bino NLSP Scenario}
As discussed before, if ${\cal O}(100)~\keV \lsim m_{3/2}\lsim {\cal O}(1)$ MeV and 
the mass splitting parameter $R$ is not so large $(R \lsim {\cal} {\cal O}(10))$,
it is possible to satisfy the cosmological constraints and
the gravitino is the candidate for the cold dark matter.
Interestingly, if $m_{3/2} \lsim {\cal O}(0.1 - 1)$ MeV,
its decay length is not much longer than the detector size.
Sizable amount of the Bino NLSP decays
occur inside of the detector.
By counting the number of these decays, 
it may be possible to determine the lifetime of the Bino, 
in other words, the gravitino mass \cite{Ishiwata:2008tp}.
This signature is peculiar compared to that of the Bino LSP scenario.

However, it is not so easy to realize such mass spectrum in the context of 
split GMSB. See Table \ref{table:summ}.

\subsubsection*{Very Light Gravitino Scenario}
In the limit of $m_{3/2} \to 0$, the gravitino abundance vanishes $\Omega_{3/2} h^2 \to 0$.
Therefore, in this case, if some sources other than the gravitino are dominant contents of the cold dark matter,
the successful cold dark matter scenario can be realized.
In this scenario, numerical study gives the upper-bound  on the gravitino mass $m_{3/2}<16$ eV \cite{Viel:2005qj}.
In this case, the NLSP decays long before the BBN era.
Therefore there are no cosmological problems other than the dark matter origin.

Table \ref{table:summ} shows that it is difficult to realize such a light gravitino scenario.
Strongly interacting GMSB models, however, have a possibility of generating sufficiently heavy sparticle masses.

In this scenario, a distinct feature is the NLSP's prompt decay into a gravitino and a SM particle.
In the case that $m_{3/2}<16~\EV$, the decay length of the NLSP is
shorter than ${\cal O}(1)$ mm for $m_{\rm NLSP}={\cal O}(100)$ GeV.

In the Bino or Wino NLSP case, the NLSP decay modes are
${\rm NLSP} \to \gamma \tilde{G}_{3/2}, Z\tilde{G}_{3/2}, \cdots$.
The emitted gravitinos cannot be detected by the detector and recognized as missing energy.
Among them, the photon signals are interesting, since there is a tiny amount of SM backgrounds \cite{Aad:2009wy}.
Therefore, it may be possible to discover the SUSY particles at low-integrated luminosity.
Even if $\sqrt{s}=7$ TeV and ${\cal L} = 1~{\rm fb}^{-1}$,
the region $M_{\tilde{g}}\lsim 900$ GeV or $M_{\tilde{W}}\lsim 500$ GeV is in reach of discovery.
A detailed analysis for this scenario will be done elsewhere.
As for the search for GMSB events with di-photon at the Tevatron,
see  Refs.~\cite{Aaltonen:2009tp,Meade:2009qv}.

One of the predictions of the split GMSB is the deformation of the gaugino masses from the GUT relation.
This can be directly tested by measuring the individual gaugino masses by using kinematical methods such as
invariant mass methods \cite{Hinchliffe:1998ys} or $M_{T2}$ methods \cite{Lester-Summers,Hamaguchi:2008hy}.
An indirect test of breaking of GUT relation at the LHC is also discussed in Ref.~\cite{Hamaguchi:2008yu}.

Next, let us discuss the case of the gluino NLSP, 
as indicated, for example, by $D$-term GMSB.
The situation is much different from the Bino or Wino NLSP case.
The gluino promptly decays into a gravitino and a gluon: 
$\tilde{g} \to g\tilde{G}_{3/2}$ inside of the detector.
Thus, typical SUSY events are di-jet and missing energy.

\subsubsection*{Wino or Gluino NLSP}
As we have discussed before, in the context of the split GMSB, 
the gaugino other than the Bino is possibly the NLSP unlike minimal GMSB.
If the Wino or gluino is the NLSP and if the gravitino is not heavy, $m_{3/2}<{\cal O}(1)~\GeV$,
the NLSP abundance is sufficiently suppressed and cosmologically safe.

First, let us consider the case of the Wino NLSP.
In this case LHC signature is similar to that of AMSB model \cite{AMSB}, 
if the neutral Wino NLSP has sizable lifetime.
Almost all of visible SUSY events are analogous to bulk scenario i.e.,
multi-jet from the gluino decay plus missing energy.
However, there is a possible difference, because of the existence of the charged Wino $\tilde{W}^{\pm}$.
Although the charged Wino has almost degenerate mass with the neutral Wino,
the charged Wino is slightly heavier by about 150 MeV from 1-loop correction
when the $\mu$ parameter is large \cite{Cheng:1998hc}.
The dominant decay mode of charged Wino is $\tilde{W}^{\pm} \to \pi^{\pm}\tilde{W}^{0}$,
and the decay length is ${\cal O}(1-10)$ cm.
This track is possibly detected by inner detectors \cite{Ibe:2006de,Asai:2008sk,Asai:2008im}.
By using this $\tilde{W}^{\pm}$ track,
various information, such as lifetime of $\tilde{W}^{\pm}$ and the masses of the Wino, Bino and gluino,
can be obtained.

A possible difference between AMSB and spit-GMSB is the relation of the three gaugino masses.
As discussed in Refs. \cite{Asai:2007sw,Asai:2008im}
in AMSB model the three gaugino masses satisfy the following relation:
\beq
\left| \frac{2g_1^2}{g_3^2}M_{\tilde{g}}
-\frac{3g_1^2}{5g_2^2}M_{\tilde{W}}
 \right|
\lsim
M_{\tilde{B}}
\lsim
\frac{2g_1^2}{g_3^2}M_{\tilde{g}}
+\frac{3g_1^2}{5g_2^2}M_{\tilde{W}},\label{eq:AMSB}
\eeq
where large radiative corrections
from the Higgs-Higgsino loops are taken into account.
In addition, in the AMSB model, it is expected that 
the Wino is a dominant component of the cold dark matter
and that
cosmic ray signals from the Wino annihilation,
which corresponds the mass of the Wino measured in collider experiments,
may be detected.

By contrast, the spit-GMSB models with the Wino NLSP need not
satisfy the relation (\ref{eq:AMSB}),
and any direct and indirect signals of the dark matter cannot be guaranteed.

Next, let us discuss the gluino NLSP.
The produced gluino forms massive hadronic state called R-hadron.
It is expected that some of R-hadrons are electrically charged and can be measured as a charged track.
Since the two gluinos are directly produced at a time, two gluino tracks have the same $p_{\rm T}$ and small velocity.
The gluino mass is determined by using information from energy loss and time of flight.
Another interesting feature of the long-lived gluino is a gluino stopping event.
Some of the R-hadrons lose their kinetic energy by interaction with matters and finally stop inside matter \cite{Arvanitaki:2005nq}.
The stopped gluinos decay into a gravitino.
By measuring such decay events, the lifetime of the gluino can be determined, 
which means determination of the gravitino mass \cite{Buchmuller:2004rq,Hamaguchi:2004df,Hamaguchi:2006vu,Asai:2009ka}.
As for the search of the stopping gluino at the Tevatron, see Ref. \cite{Abazov:2007ht}.

\subsubsection*{Small $\mu$ Parameter Scenario}
We have seen in Eq.~(\ref{eq:muorder})
that the $\mu$ parameter is the same order as the sfermion masses.
However, it may be also possible to 
get small $\mu$ parameter by tuning model parameters.
This is because in the split GMSB models,
the running of RG equation is possibly short and
in some case, colored SUSY particles are relatively light, compared to the minimal GMSB case.
The small $\mu$ parameter reduces the Bino-like NLSP abundance and
relaxes the cosmological constraints.
Therefore if the $\mu$ parameter is small enough,
the Bino-like NLSP scenario is possible.
In an extreme case,
if the $\mu$ parameter is so small that the NLSP becomes dominantly Higgsino,
the NLSP abundance is given as
\beq
\Omega_{\rm Higgsino} h^2 \simeq 10^{-3}\left(\frac{\mu}{100~\GeV}\right)^2.
\eeq
In this case, the cosmological constraints become very weak like the Wino NLSP case.

If the $\mu$ parameter is small enough, the situation is similar to the split SUSY \cite{splitSUSY} or focus-point \cite{FocusPoint}, except for the differences in the nature of the dark matter.
In these cases, the dark matter is possibly the neutralino.
If the dark matter is the neutralino,
the properties of the dark matter can be measured from LHC signature.
As for the dark matter measurements in sfermion decoupling limit, see Refs. \cite{Baltz:2006fm,White:2010jp}.
In the case of the neutralino dark matter, the dark matter abundance reproduced from the LHC measurements are
expected to be the present dark matter abundance $\Omega h^2=0.1$.
On the other hand, in the context of GMSB, it is not necessary to reproduce the correct dark matter abundance.
Instead, the reproduced dark matter abundance must be $\Omega h^2 \ll{0.1}$ to evade the BBN constraint, assuming
the lifetime of the neutralino is longer than about one second.

\section{Discussion and Conclusion}\label{sec:6}

In this paper, we discuss that in a wide class of non-minimal GMSB models
it is difficult to generate the gaugino masses at leading order ${\cal O}(F/M)$.
The absence of the leading term leads to a very large mass splitting between the masses of the SSM gauginos and sfermions, and also leads to the
violation of the GUT relation of the gaugino masses.
One may think that such mass splitting is not desirable from the viewpoints of gauge hierarchy problem and 
GUT unification, but there seems to be no objective criterion for the naturalness as discussed in the Introduction.

In this work, we have found that such a mass splitting is not favorable from the cosmological viewpoint
if the Bino is the NLSP. We discuss, however, that the NLSP other than the Bino is plausible in non-minimal GMSB models, and
the cosmological constraint becomes much weaker in that case.
Then we discuss the LHC signature of split GMSB models which evade cosmological constraints.
We have seen that
many cases satisfying the cosmological constraints 
produce peculiar events at the LHC, for example in flight decay of the NLSP or R-hadrons.  
Among them, the gluino NLSP scenario is interesting, since the gluino tends to be light in the split GMSB models.

Let us comment on the case that the gravitino is heavy and not the LSP.
In the usual GMSB model, it is undesirable that the gravitino gets heavy mass, since gravity mediation contribution to the sfermion masses become non-negligible and it may cause flavor-violating neutral current
through Planck-suppressed higher dimensional operators.
However, if the scalar particles are heavy enough,
this is not always the case, because such processes are suppressed 
by heavy scalar masses.
Thus, it may be possible that the gravitino is not the LSP and that
the MSSM LSP is the cold dark matter in the context of split GMSB (see Ref.~\cite{Ibe:2009pq} for a recent discussion).
If the Bino is the LSP, it seems to be difficult to realize the correct dark matter abundance, since the Bino dark matter easily over-close the Universe in the split GMSB models as seen before. If the Wino is the LSP and a dominant component of the dark matter, 
the mass of the Wino is about 3 TeV in order to explain the present dark matter abundance. Unfortunately, this makes it difficult to discover the SUSY particles at the LHC.
\\
\\
{\bf Note added:} After the completion of this work we received a paper~\cite{Dumitrescu:2010ha},
which has some overlap with our paper.

\section*{Acknowledgements}
S.~S. and K.~Y. would like to thank E.~Nakamura, K.~Hamaguchi, R.~Sato and T.~T.~Yanagida for useful discussions.
This work is supported in part by JSPS
Research Fellowships for Young Scientists and by
World Premier International Research Center Initiative, MEXT, Japan.
M.~Y. is also supported in part by DOE grant DE-FG03-92-ER40701 and Global COE Program for Physical Sciences Frontier at 
the University of Tokyo, MEXT, Japan.

\appendix 
\setcounter{equation}{0}
\renewcommand{\theequation}{\Alph{section}.\arabic{equation}}

\section{MSSM Mass Spectrum in Semi-Direct Gauge Mediation} \label{app:A}
In this Appendix we compute the MSSM mass spectrum in semi-direct gauge mediation.
Semi-direct gauge mediation models can be characterized as follows.
There are messenger fields $\P,~\tilde{\P}$ charged under both the SM gauge group $SU(3)_C \times SU(2)_L \times U(1)_Y$
and a gauge group $G_h$, which couples to a SUSY breaking hidden sector. $\P,~\tilde{\P}$ only have a bare mass term $W=M\P \tilde{\P}$ 
in the superpotential,
and the SUSY breaking is mediated to the MSSM sector by loops involving $G_h$ gauge fields and the messengers.
For example, in the semi-direct gauge mediation model discussed in section~\ref{sec:semimi}, $\P=l,\tilde{\P}=\tilde{l}$ and $G_h=SU(2)_{\rm hid}$.

In Ref.~\cite{Argurio:2009ge}, gaugino and sfermion mass spectrum is studied by explicit diagram calculations under the assumption
that the gauge group $G_h$ is not broken (or broken at a negligibly low scale). Here we calculate the soft masses in the opposite limit; i.e.
the breaking scale of $G_h$ is very high. More precisely, we assume
that the $G_h$ breaking scale $M_V$ is much larger than the SUSY breaking scale and the messenger mass scale.

After integrating out the $G_h$ gauge degrees of freedom at the scale $M_V$, we obtain higher dimensional operators which couple 
the messenger fields and the hidden sector fields in the effective K\"ahler potential (see e.g. Eqs.~(\ref{eq:0loop}, \ref{eq:1loop})).
Then, neglecting the hidden sector dynamics~\footnote{
Sometimes hidden sector dynamics becomes important when we consider RG evolution. 
} and simply replacing the hidden sector fields by their VEVs, we obtain the following
``softly broken messenger Lagrangian'',
\beq
{\mathcal L} =\sum_r \left[ \int  \! d^4\h\, (\Z_{r+}\P_r^\dagger \P_r+ \Z_{r-}\tilde{\P}_r^\dagger \tilde{\P}_r)+\int\! d^2\h\, M_r \P_r \tilde{\P}_r+{\rm h.c.} \right], 
\eeq
where $\P_r,~\tilde{\P}_r$ are the messenger fields in the representation $r$ and $\bar{r}$ of $SU(3)\times SU(2)\times U(1) $,
$M_r$ are SUSY invariant mass of the messengers, and $\Z_{r\pm}$ are SUSY breaking spurion fields.
We assume that $\Z_{r\pm}$ are of the form,
\beq
\log \Z_{r\pm} = \log Z_{r\pm}+\frac{1}{2}(f_r\h^2+{\rm h.c.}) +(\mp D_r+d_r)\h^2\bar{\h^2}, \label{eq:messwave}
\eeq
where $Z_{r\pm}$ are usual wave function renormalizations, and $D_r,~d_r$ and $f_r$ are soft terms of order 
\beq
D_r\sim v^2,~~~d_r\sim \frac{\a_h}{4\pi} v^2,~~~f_r\sim \frac{\a_h}{4\pi} v,
\eeq
where $\a_h$ is the fine structure constant of $G_h$, and $v$ is some mass scale.
The meaning of this assumption is as follows.
$D_r$ is a tree level contribution to the messenger mass spectrum, which maintains the messenger supertrace being zero.
$d_r$ and $f_r$ are quantum corrections to the messenger mass spectrum, which does not necessarily obey the supertrace condition.
The semi-direct mediation model discussed in section~\ref{sec:semimi} satisfies this assumption.

In MSSM phenomenology,  one takes ``softly broken MSSM Lagrangian'' as their starting point and calculate the low energy consequences.
Analogously, our approach here is to consider ``softly broken messenger + MSSM Lagrangian'' as a starting point and then integrate out the messengers to obtain
the MSSM mass spectrum. The softly broken Lagrangian at the scale $M_V$ is given by,
\beq
{\cal L} &=& \sum_r \left[ \int \! d^4\h\, (\Z_{r+}\P_r^\dagger \P_r+ \Z_{r-}\tilde{\P}_r^\dagger \tilde{\P}_r)+\int \! d^2 \h \, M_r \P_r \tilde{\P}_r+{\rm h.c.} \right] \nonumber \\
&&+\sum_i \int \! d^4\h\, \Z_i \f^\dagger_i \f_i +\sum_{a=1,2,3} \int \!d^2\h \, \frac{1}{4g^2_a}W^2_a+{\rm h.c.}, \label{eq:softlagrangian}
\eeq
where $\f_i$ are the MSSM matter fields, and $W_a$ ($a=1,2,3$) the field strength of $U(1)_Y,SU(2)_L$ and $SU(3)_C$, respectively
We have omitted the MSSM superpotential for simplicity.
$\Z_i$ give soft terms to the MSSM fields, and are given by
\beq
\log \Z_i=\log Z_i-(a_{\f_i}\h^2+{\rm h.c.}) -m^2_{\f_i}\h^2\bar{\h^2}, \label{eq:mssmwave}
\eeq
where $m^2_{\f_i}$ is the soft mass of the sfermion in $\f_i$.
$g_a$ are holomorphic gauge couplings. As discussed in section~\ref{sec:3}, in semi-direct gauge mediation
$1/g^2_a$ have no $\h^2$ terms which are related to the leading order gaugino mass.
On the other hand, there is no reason that $a_{\f_i}$ and $m^2_{\f_i}$ are exactly zero at the scale $M_V$. However, from simple loop counting, these are at most of order
\beq
a_{\f_i} &\sim& \left(\frac{\a_v}{4\pi}\right)^2f_r \sim
 \frac{\a_h}{4\pi}\left(\frac{\a_v}{4\pi}\right)^2v, \\
 m^2_{\f_i} &\sim& \left(\frac{\a_v}{4\pi}\right)^2 d_r \sim
 \frac{\a_h}{4\pi}\left(\frac{\a_v}{4\pi}\right)^2v^2, \label{eq:iniorder}
\eeq
that is, $a_{\f_i}$ and $m^2_{\f_i}$ are 2-loop suppressed by the SM couplings $\a_v$ 
compared with $f_r$ and $d_r$, respectively. $m^2_{\f_i}$ does not receive a contribution of order $(\a_v^2/4\pi)^2D_r$, since
$D_r$ in the messenger soft mass satisfies the supertrace condition, and the effect of $D_r$ on the MSSM soft mass can be definitely calculable, as we will see.
Furthermore, $a_{\f_i}$ and $m^2_{\f_i}$ have no logarithmic enhancement at the scale $M_V$, since we have assumed that $M_V$ is the highest energy
scale of the model. The absence of logarithmic enhancement at the scale $M_V$ will be important for the calculation of $a_{\f_i}$ and $m^2_{\f_i}$ at the messenger mass scale.

Let us first calculate the gaugino masses. For that, it is quite important to use a regularization scheme which manifestly maintains the spurious SUSY
which acts on the couplings $\Z_{r\pm},\Z_i$ and $g_a$ seen as superfields~\cite{ArkaniHamed:1998kj}.
One of the easiest ways to regularize the messenger loops at 1-loop is to use the Pauli-Villars regularization.
We introduce regulator fields $\W_r$ and $\tilde{\W}_r$ which have opposite statistics to the messenger fields $\P_r$ and $\tilde{\P}_r$.
We take the regulator Lagrangian~\cite{ArkaniHamed:1998kj},
\beq
\sum_r \left[ \int \! d^4\h\, (\Z_{r+}\W_r^\dagger \W_r+ \Z_{r-}\tilde{\W}_r^\dagger \tilde{\W}_r)+\int \!d^2\h \,\L_r \W_r \tilde{\W}_r+{\rm h.c.} \right] ,
\eeq
where $\L_r$ are regulator masses. In this form, the spurious SUSY is manifestly maintained by the regularization. After ensuring the manifest SUSY, we can
calculate the gaugino masses by using component field diagrams (i.e. not necessarily supergraphs). Such a calculation was done in Ref.~\cite{Poppitz:1996xw}.
After rescaling the fields such that $Z_{r\pm}=Z_i=1$, the gaugino masses are given by
\beq
M_a =  \frac{g_a^2}{16\pi^2} \sum_r 2t_a(r) \left[ G(M_r,f_r,D_r,d_r)-G(\L_r,f_r,D_r,d_r)\right] , \label{eq:gmassformula}
\eeq
where $t_a(r)$ are the Dynkin index of the representation $r$, and $G(M_r,f_r,D_r,d_r)$ is a function whose explicit form is given in Ref.~\cite{Poppitz:1996xw}.
The first term is the contribution from the messenger loops, and the second term from the regulator loops.
$G(M_r,f_r,D_r,d_r)$ can be expanded as
\beq
G(M_r,f_r,D_r,d_r)= f_r \left( 1+\frac{2d_r}{3M_r^2}+\frac{|f_r|^2}{6M_r^2}+\frac{D_r^2}{6M_r^4} +\cdots \right). \label{eq:messloop}
\eeq
Eq.~(\ref{eq:gmassformula}) is finite in the limit $\L_r \to \infty$. 
This is consistent, because if some divergence had appeared at this order, we could not cancel the divergence because we have shown that $1/g_a^2$ have no $\h^2$
term and hence we cannot have a counterterm to cancel the divergence. 

An important point in the introduction of the regulator fields is that the regulator contribution is in fact non-vanishing in the limit $\L_r \to \infty$,
\beq
\lim_{\L_r \to \infty} G(\L_r,f_r,D_r,d_r)=f_r,
\eeq
and this cancels the leading term of Eq.~(\ref{eq:messloop}). Thus, in this effective field theory calculation,
 we have shown the vanishing of the leading term of the gaugino masses discussed in
section~\ref{sec:2.1}. The final result for the gaugino masses is (assuming $|f_r|^2/d_r \sim \a_h/4\pi \ll 1$),
\beq
M_a &=&  \frac{g_a^2}{16\pi^2} \sum_r 2t_a(r) f_r \left( \frac{2d_r}{3M_r^2}+\frac{D_r^2}{6M_r^4} +\cdots \right),\label{eq:gauginoresult}
\eeq
where dots denote terms which are suppressed compared with the explicitly written terms if $\a_h/4\pi \ll 1$ and $v/M_r \ll 1$.

In the above calculation we used the Pauli-Villars regularization. But a more conventional regularization scheme in SUSY theory is the dimensional reduction
with modified minimal subtraction ($\overline{\rm DR}$). Now we calculate the gaugino masses in this scheme to see the consistency of the calculation. 
We have to take care when we use the $\overline{\rm DR}$ scheme with component field calculation
as is practically done, 
since that calculation does not maintain the manifest superfield structure. 
In Ref.~\cite{ArkaniHamed:1998kj}, the real gauge coupling superfield $R_a$ is defined, given by
\beq
R_a={\rm Re}(g_{a}^{-2})+\frac{t_a(A)}{8\pi^2}\log[{{\rm Re} (g_a^{-2})}]-\sum_{s=\pm,r} \frac{t_a(r)}{8 \pi^2} \log\Z_{r,s}
-\sum_{i} \frac{t_a(i)}{8 \pi^2} \log\Z_{i} +{\cal O}(g^2_a).  \nonumber \\ \label{eq:realcouplingsufi}
\eeq
Then it was shown that the gauge coupling constant $g_{Ra}$ and the gaugino masses $M_a$ defined in the $\overline{\rm DR}$ scheme should 
be related to the component of $R_a$ as
\beq
R_a=\frac{1}{g_{Ra}^{2}}+\left(\frac{M_a}{g_{Ra}^2}\h^2+{\rm h.c.}\right)+{\cal O}(\h^2\bar{\h}^2). \label{eq:comprealcoup}
\eeq
As is well known, the holomorphic gauge coupling $g_a$ runs only at 1-loop level. On the other hand, the real gauge coupling $g_{Ra}$ runs
at all orders in perturbation theory, and is interpreted as a physical gauge coupling. Eq.~(\ref{eq:realcouplingsufi}) originally appeared in 
Ref.~\cite{Shifman:1986zi} in the supersymmetric limit and then extended in Ref.~\cite{ArkaniHamed:1998kj} as a superfield equation with soft terms included. 

Using Eqs.~(\ref{eq:messwave}), (\ref{eq:realcouplingsufi}) and (\ref{eq:comprealcoup}), we see that the gaugino masses are 
non-vanishing already at the tree level,
\beq
M^{\rm tree}_a=-\frac{g_a^2}{16\pi^2}\sum_r 2t_a(r)f_r+\cdots, \label{eq:gauginotree}
\eeq
even before integrating out the messengers. Here dots denote higher order terms in $g_a$.
Then, integrating out the messengers, we obtain the one-loop contribution
\beq
M^{\rm 1-loop}_a =  \frac{g_a^2}{16\pi^2} \sum_r 2t_a(r)  G(M_r,f_r,D_r,d_r). \label{eq:gauginoloop} 
\eeq
Adding Eqs.~(\ref{eq:gauginotree}) and (\ref{eq:gauginoloop}), we obtain the same result (\ref{eq:gauginoresult}) as in the case of the Pauli-Villars regularization.

Let us next consider the sfermion masses. In this paper, we assume that there are no dangerous contributions to the $D$-term of $U(1)_Y$
from messenger loops. See Ref.~\cite{Dimopoulos:1996ig} for a mechanism, called messenger parity, to cancel such contributions. The model discussed in
section~\ref{sec:semimi} has an approximate messenger parity~\cite{Seiberg:2008qj}. In the following, we only consider 2-loop contributions to the sfermion masses
which are not related to the $U(1)_Y$ $D$-term.

It is known~\cite{Poppitz:1996xw} that there is a divergence in the loop contribution to the sfermion masses 
due to the non-vanishing of the messenger mass supertrace. 
This divergence suggests that we need to renormalize the sfermion masses.
The renormalization group (RG) equation for the ``MSSM + messenger'' sector is similar to the one in the MSSM (see Ref.~\cite{Martin:1997ns}).
Here we derive the RG equations using the technique of Ref.~\cite{ArkaniHamed:1998kj}.

In a supersymmetric limit, the RG equation for the wave function renormalization $Z_i$ at the 1-loop level is given by
\beq
\frac{\partial \log Z_i}{\partial \log\mu}=-\g_i= \sum_a C_a(i)\frac{g_a^2}{4\pi^2},
\eeq
where $\mu$ is a renormalization scale, $\g_i$ the anomalous dimension of $\f_i$, and $C_a(i)$ the quadratic Casimir for $\f_i$. 
We have neglected the MSSM Yukawa couplings for simplicity.
Then we analytically continue~\cite{ArkaniHamed:1998kj} the equation into superspace,
\beq
\frac{\partial \log \Z_i}{\partial \log\mu}= \sum_a C_a(i)\frac{R_a^{-1}}{4\pi^2},
\eeq
where $R_a$ is defined by Eq.~(\ref{eq:realcouplingsufi}). Taking only the most important parts, we obtain
\beq
\frac{\partial \log \Z_i}{\partial \log\mu} \simeq  \sum_a \left[ C_a(i)\frac{g_a^2}{4\pi^2}+\sum_{s=\pm,r}8C_a(i)t_a(r)\left(\frac{g_a^2}{16\pi^2}\right)^2
\log \Z_{r,s}\right].
\eeq
Taking the $\h^2\bar{\h}^2$ and $\h^2$ component of this equation and integrating from $M_V$ to the messenger mass scale, we obtain $m^2_{\f_i}$ and
$a_{\f_i}$ at the messenger mass scale,
\beq
m^2_{\f_i}|_{\rm mess} &=&
m^2_{\f_i}|_{\mu = M_V} +\sum_{a} \sum_r \left(\frac{g_a^2}{16\pi^2} \right)^2 C_a(i) t_a(r)  8d_r \log \frac{M_V^2}{M_r^2} +{\cal O}(g_a^6),\label{eq:rgterm} \\
a_{\f_i}|_{\rm mess} &=& a_{\f_i}|_{\mu = M_V}+\sum_{a} \sum_r \left(\frac{g_a^2}{16\pi^2} \right)^2 C_a(i) t_a(r)  4f_r \log \frac{M_V^2}{M_r^2} +{\cal O}(g_a^6).\label{eq:Aterminsemi}
\eeq
If we assume that $\log(M_V/M_r)$ are large, the initial value $m^2_{\f_i}|_{M_V}$ and $a_{\f_i}|_{M_V}$ 
of order Eq.~(\ref{eq:iniorder}) is small compared with the log enhanced RG contribution.

There is one more important contribution at the threshold of the messengers. To see the most important contribution, let us take the limit $\a_h \to 0$.
Then, the supertrace of the messenger mass matrix is zero, and 
\beq
\Z_{r\pm}=\mp D_r \h^2\bar{\h}^2.
\eeq
This is the same as in the $D$-term gauge mediation~\cite{Nakayama:2007cf}, and the messenger threshold correction is given by
\beq
-\sum_{a} \sum_r  \left(\frac{g_a^2}{16\pi^2} \right)^2 C_a(i) t_a(r)\left(\frac{7D_r^4}{9M_r^6} +{\cal O}(D_r^6/M_r^{10})\right). \label{eq:dterm}
\eeq
The sum of Eqs.~(\ref{eq:rgterm}) and (\ref{eq:dterm}) gives the leading part of the sfermion masses.

Let us comment on $a_{\f_i}$. By a rescaling $\f_i \to \exp(a_{\f_i}\h^2)\f_i$, we can eliminate $a_{\f_i}$ from the K\"ahler potential.
Then, the MSSM superpotential $W=\sum y_{ijk}\f_i\f_j\f_k+\sum \m_{ij} \f_i \f_j$ becomes 
\beq
W \to \sum (1+a_{\f_i}\h^2+a_{\f_j}\h^2+a_{\f_k}\h^2)y_{ijk}\f_i\f_j\f_k+\sum (1+a_{\f_i}\h^2+a_{\f_j}\h^2)\m_{ij} \f_i \f_j,
\eeq
and this leads to the $A$ term $a_{ijk}$ and the $B$ term $b_{ij}$,
\beq
a_{ijk}&=&(a_{\f_i}+a_{\f_j}+a_{\f_k})y_{ijk}, \\
b_{ij}&=&(a_{\f_i}+a_{\f_j})\m_{ij},
\eeq
which are 2-loop effects in the SM gauge couplings and are small. 
Furthermore, from Eqs.~(\ref{eq:mssmwave}), (\ref{eq:realcouplingsufi}) and (\ref{eq:comprealcoup}) (or equivalently, 
from the rescaling anomaly~\cite{Konishi:1983hf}
in the transformation $\f_i \to \exp(a_{\f_i}\h^2)\f_i$) we get a contribution to the gaugino masses
\beq
M_a^{\rm 3-loop}= \frac{g^2_a}{16 \pi^2}\sum_{i}2t_a(i) a_{\f_i}. \label{eq:threeloopinsemi}
\eeq
This is the contribution discussed below Eq.~(\ref{eq:nonholoop}) in section~\ref{sec:2}.
$M_a^{\rm 3-loop}$ is a 3-loop effect in the SM gauge couplings as expected. 
But note that this contribution is not suppressed by $v/M_r$ and may become important
if $v/M_r$ is too small. We neglect $M_a^{\rm 3-loop}$ in this paper.

In summary, the soft masses are given by
\beq
M_a &\simeq&  \frac{g_a^2}{16\pi^2} \sum_r 2t_a(r) f_r \left( \frac{2d_r}{3M_r^2}+\frac{D_r^2}{6M_r^4} \right), \label{eq:massformula1}\\
m_{\f_i}^2 &\simeq& \sum_{a=1,2,3} \sum_r \left(\frac{g_a^2}{16\pi^2} \right)^2 C_a(i) t_a(r) \left( -\frac{7D_r^4}{9M_r^6} + 8d_r \log \frac{M_V^2}{M_r^2}  \right),
\label{eq:massformula2}
\eeq
where we have neglected all terms which are small in the limit $\a_h/4\pi \ll 1$, $v \ll M_r$ and $\log (M_V/M_r) \gg 1$.

\end{document}